\documentclass[superscriptaddress,twocolumn,prx,floatfix,nofootinbib]{revtex4-1}

\usepackage{graphicx}
\usepackage{bm}
\usepackage{amsmath}
\usepackage{hyperref} \hypersetup{colorlinks=true,citecolor=blue,linkcolor=blue,urlcolor=blue}
\usepackage{bbm}
\usepackage[export]{adjustbox}
\usepackage[dvipsnames]{xcolor}
\usepackage{multirow}
\usepackage{array} 

\bibpunct{[}{]}{,}{n}{}{} 

\def\AGL{{\small AGL}}

\def\AEL{{\small AEL}}
\def\APL{{\small APL}}
\def\QHAAPL{{\small QHA-APL}}
\def\AAPL{{\small AAPL}}
\def\AFLOW{{\small AFLOW}}
\def\AFLOWPOCC{{\small AFLOW-POCC}}
\def\AFLUX{{\small AFLUX}}
\def\AFLOWML{{\small AFLOW-ML}}
\def\AFLOWHULL{{\small AFLOW-HULL}}
\def\LUX{{\small LUX}}
\def\URI{{\small URI}}
\def\JSON{{\small JSON}}

\def\AFLOWpi{{\small AFLOW$\pi$}}
\def\PAOFLOW{{\small PAOFLOW}}
\def\VASP{{\small VASP}}
\def\ICSD{{\small ICSD}}
\def\RESTAPI{{\small {\small REST-API}}}
\def\GIBBS{{\small GIBBS}}

\def\AFLOWSYM{{\small AFLOW-SYM}} 
\def\QUANTUMESPRESSO{\textsc{Quantum {\small ESPRESSO}}} 
\def\FHIAIMS{{\small FHI-AIMS}} 
\def\ABINIT{{\small ABINIT}} 

\def\sDebye{{\substack{\scalebox{0.6}{D}}}}
\def\sD{\sDebye}
\def\sDFT{{\substack{\scalebox{0.6}{DFT}}}}

\def\sVRH{{\substack{\scalebox{0.6}{VRH}}}}

\def\svib{{\substack{\scalebox{0.6}{vib}}}}

\def\sS{{\substack{\scalebox{0.6}{S}}}}

\def\sV{{\substack{\scalebox{0.6}{V}}}}
\def\sB{{\substack{\scalebox{0.6}{B}}}}

\usepackage{soul}
\setstcolor{red} 

\makeatletter \renewcommand\frontmatter@abstractwidth{\dimexpr\textwidth\relax} \makeatother 

\begin{document}

\title{\LARGE The AFLOW Fleet for Materials Discovery}

\def\MEMS{{Department of Mechanical Engineering and Materials Science, Duke University, Durham, NC 27708, USA}}
\def\CMS{{Center for Materials Genomics, Duke University, Durham, NC 27708, USA}}
\def\FHI{{Fritz-Haber-Institut der Max-Planck-Gesellschaft, 14195 Berlin-Dahlem, Germany}}
\def\DUKE{{Materials Science, Electrical Engineering, Physics and Chemistry, Duke University, Durham, NC 27708, USA}} 
\def\CMICH{{Department of Physics and Science of Advanced Materials Program, Central Michigan University, Mount Pleasant, MI 48859, USA}}
\def\CMICHS{{Science of Advanced Materials Program, Central Michigan University, Mount Pleasant MI, USA}}
\def\UNCa{{Laboratory for Molecular Modeling, Division of Chemical Biology and Medicinal Chemistry, UNC}}
\def\UNCb{{Eshelman School of  Pharmacy, University of North Carolina, Chapel Hill, NC 27599, USA}}
\def\CEA{{CEA, LITEN, 17 Rue des Martyrs, 38054 Grenoble, France}}
\def\UNT{{Department of Physics and Department of Chemistry, University of North Texas, Denton, TX 76203, USA}}
\def\CNRMODENA{{CNR-NANO Research Center S3, 41125 Modena, Italy}}
\def\NRCN{{Department of Physics, NRCN, P.O. Box 9001, Beer-Sheva 84190, Israel}}


\author{
  Cormac~Toher$^{1,2,\star}$,
Corey~Oses$^{1,2}$,
David~Hicks$^{1,2}$,
Eric~Gossett$^{1,2}$,
Frisco~Rose$^{1,2}$,
Pinku~Nath$^{1,2}$,
Demet~Usanmaz$^{1,2}$,
Denise~C.~Ford$^{1,2}$,
Eric~Perim$^{1,2}$,
Camilo~E.~Calderon$^{1,2}$,
Jose~J.~Plata$^{1,2,3}$
Yoav~Lederer$^{1,2,4}$,
Michal~Jahn\'{a}tek$^{1,2}$,
Wahyu~Setyawan$^{1,2}$,
Shidong~Wang$^{1,2}$,
Junkai~Xue$^{1,2}$,
Kevin~Rasch$^{1,2}$,
Roman~V.~Chepulskii$^{1,2}$,
Richard~H.~Taylor$^{1,2,5}$
Geena~Gomez$^{1,2}$,
Harvey~Shi$^{2}$,
Andrew~R.~Supka$^{6,7}$,
Rabih~Al~Rahal~Al~Orabi$^{6,8}$,
Priya~Gopal$^{6}$,
Frank~T.~Cerasoli$^{9}$,
Laalitha~Liyanage$^{9}$,
Haihang~Wang$^{9}$,
Ilaria~Siloi$^{9}$,
Luis~A.~Agapito$^{9}$,
Chandramouli~Nyshadham$^{10}$,
Gus~L.~W~Hart$^{10}$,
Jes{\'u}s~Carrete$^{11}$,
Fleur~Legrain$^{12,13}$,
Natalio~Mingo$^{13}$,
Eva~Zurek$^{14}$,
Olexandr~Isayev$^{15,16}$,
Alexander~Tropsha$^{15,16}$,
Stefano~Sanvito$^{17,2}$,
Robert~M.~Hanson$^{18}$,
Ichiro~Takeuchi$^{19,20}$,
Michael~J.~Mehl$^{21,2}$,
Aleksey~N.~Kolmogorov$^{22,2}$,
Kesong~Yang$^{23,2}$,
Pino~D'Amico$^{24,25}$,
Arrigo~Calzolari$^{24,2,9}$,
Marcio~Costa$^{26}$,
Riccardo~De~Gennaro$^{27}$,
Marco~Buongiorno~Nardelli$^{9,2}$,
Marco~Fornari$^{6,7,2}$,
Ohad~Levy$^{1,2,4}$,
and Stefano~Curtarolo$^{1,2,28,\dagger}$
}

\begin{abstract}
\noindent{\footnotesize \it
$^{1}$\MEMS;
$^{2}$\CMS;
$^{3}$Departamento de Qu\'{i}mica F\'{i}sica, Universidad de Sevilla, 41012 Sevilla, Spain;
$^{4}$\NRCN;
$^{5}$Department of Materials Science and Engineering, Massachusetts Institute of Technology, MA 02139, USA;
$^{6}$\CMICH;
$^{7}$\CMICHS;
$^{8}$Solvay, Design and Development of Functional Materials Department, AXEL'ONE Collaborative Platform - Innovative Materials, 69192 Saint Fons Cedex, France;
$^{9}$\UNT;
$^{10}$Department of Physics and Astronomy, Brigham Young University, Provo, UT 84602, USA;
$^{11}$Institute of Materials Chemistry, TU Wien, A-1060 Vienna, Austria;
$^{12}$Universite ́e Grenoble Alpes, 38000 Grenoble, France;
$^{13}$\CEA;
$^{14}$Department of Chemistry, State University of New York at Buffalo, Buffalo, NY 14260, USA;
$^{15}$\UNCa;
$^{16}$\UNCb;
$^{17}$School of Physics, AMBER and CRANN Institute, Trinity College, Dublin 2, Ireland;
$^{18}$Department of Chemistry, St. Olaf College, Northfield, MN 55057, USA;
$^{19}$Center for Nanophysics and Advanced Materials,vUniversity of Maryland, College Park, MD 20742, USA;
$^{20}$Department of Materials Science and Engineering, University of Maryland, College Park, MD 20742, USA;
$^{21}$United States Naval Academy, Annapolis, MD 21402, USA;
$^{22}$Department of Physics, Binghamton University, State University of New York, Binghamton, NY 13902, USA;
$^{23}$Department of NanoEngineering, University of California San Diego, La Jolla, CA 92093-0448, USA;
$^{24}$\CNRMODENA;
$^{25}$Dipartimento di Fisica, Informatica e Matematica, Universita ́ di Modena and Reggio Emilia, 41125 Modena, Italy;
$^{26}$Brazilian Nanotechnology National Laboratory (LNNano), CNPEM, 13083-970 Campinas, Brazil;
$^{27}$Dipartimento di Fisica, Universit`a di Roma Tor Vergata, 00133 Roma, Italy.
$^{28}$\FHI } \ \\

\noindent
The traditional paradigm for materials discovery has been recently expanded to incorporate substantial data driven research.
With the intent to accelerate the development and the deployment of new technologies, the \AFLOW\ Fleet for computational
materials design automates high-throughput first principles calculations, and provides tools for data verification and
dissemination for a broad community of users.
\AFLOW\ incorporates different computational modules to robustly determine thermodynamic stability, electronic band structures,
vibrational dispersions, thermo-mechanical properties and more.
The \AFLOW\ data repository is publicly accessible online at \verb|aflow.org|, with more than 1.7 million materials entries and
a panoply of queryable computed properties. Tools to programmatically search and process the data, as well as to perform online machine
learning predictions, are also available.
\end{abstract}


\maketitle
\section{Introduction}

The \AFLOW\ Fleet is an integrated software infrastructure for automated materials design~\cite{nmatHT}
centered around the \underline{A}utomatic \underline{Flow} (\AFLOW)~\cite{curtarolo:art65} framework for computational materials science.
It features multiple scientific software packages, including the \AFLOW\ high-throughput framework, 
the \AFLOW$\pi$ \cite{curtarolo:art127} medium-throughput framework, and the {\small PAOFLOW} \cite{paoflow} utility for electronic structure analysis, 
along with the \AFLOW.org data repository~\cite{curtarolo:art75}, its associated \underline{re}presentational \underline{s}tate \underline{t}ransfer 
\underline{a}pplication \underline{p}rogramming \underline{i}nterface ({\small REST-API}) \cite{curtarolo:art92}, and the \AFLUX\ Search-{\small API} \cite{curtarolo:art128}.
These elements are well integrated with one another: a Python+\JSON\ (\underline{J}ava\underline{S}cript \underline{O}bject \underline{N}otation) interface 
connects \AFLOW, \AFLOW$\pi$ and {\small PAOFLOW}; and all software packages access the \AFLOW.org repository via the {\small REST-API} and the Search-{\small API}.

Similar infrastructure has been developed by initiatives such as the Materials Project \cite{APL_Mater_Jain2013}, NoMaD \cite{nomad}, OQMD \cite{oqmd.org}, 
the Computational Materials Repository \cite{cmr_repository}, and AiiDA \cite{Pizzi_AiiDA_2016}.  
The Materials Project uses the \verb|pymatgen| \cite{CMS_Ong2012b} Python-language data generation software infrastructure, and their repository is available at \verb|materialsproject.org|.
The \underline{No}vel \underline{Ma}terials \underline{D}iscovery (NoMaD) Laboratory maintain an aggregate repository available at \verb|nomad-repository.eu|, incorporating 
data generated by other frameworks including \AFLOW. 
The \underline{O}pen \underline{Q}uantum \underline{M}aterials \underline{D}atabase (OQMD) \cite{oqmd.org} uses tools such as {\tt qmpy} to generate their database, which can be accessed at \verb|oqmd.org|.
The \underline{A}tomic \underline{S}imulation \underline{E}nvironment (ASE) \cite{ase} is used to generate the Computational Materials Repository, available at \verb|cmr.fysik.dtu.dk|. 
The ASE scripting interface is also used by the \underline{A}utomated \underline{I}nteractive \underline{I}nfrastructure and \underline{Da}tabase (AiiDA) framework available at \verb|aiida.net|, 
which revolves around relational databases for its overall design and data storage.
Additional materials design utilities include the \underline{H}igh-\underline{T}ool\underline{k}it (HTTK) and the associated Open Materials Database, \verb|httk.openmaterialsdb.se|,
as well as the Materials Mine database available at \verb|www.materials-mine.com|, while computationally predicted crystal structures can be obtained from the Theoretical Crystallography Open Database 
at \verb|www.crystallography.net/tcod/|.

The \AFLOW\ Fleet employs \underline{d}ensity \underline{f}unctional \underline{t}heory (DFT) to perform the quantum mechanical calculations required to obtain materials properties from first principles. 
These DFT calculations are carried out by external software packages, namely the \underline{V}ienna \underline{A}b initio \underline{S}imulation \underline{P}ackage (\VASP) \cite{kresse_vasp, vasp} in the case of \AFLOW, and \QUANTUMESPRESSO\ \cite{qe, Giannozzi:2017io} in the case of \AFLOW$\pi$.
Results are stored in the \AFLOW.org repository~\cite{curtarolo:art75} and made freely available online via the \verb|aflow.org| web portal, 
which is programmatically accessible and searchable via the \AFLOW\ Data {\small REST-API}~\cite{curtarolo:art92} and \AFLUX\ Search-{\small API} \cite{curtarolo:art128} respectively.
The repository currently contains calculated properties for over 1.7 million materials entries, including both experimentally observed and theoretically predicted structures, and new results are continuously being added. 
This \AFLOW\ data is successfully applied to
(\textit{i}) formulate descriptors for the formation of disordered materials such as metallic glasses~\cite{curtarolo:art112}, 
(\textit{ii}) find new magnetic materials~\cite{curtarolo:art109} and superalloys \cite{curtarolo:art113}, 
(\textit{iii}) generate phase diagrams for alloy systems \cite{curtarolo:art106, curtarolo:art117, Lederer_HEA_2017} and identify new ordered compounds \cite{curtarolo:art49,curtarolo:art51, curtarolo:art53, curtarolo:art126}, and 
(\textit{iv}) train machine learning models to identify potential superconductors \cite{curtarolo:art94} and 
predict electronic and thermo-mechanical properties \cite{curtarolo:art124}.

\section{AFLOW: Efficient data generation}

The \AFLOW\ framework for computational materials science automates 
the full workflow for materials properties calculations~\cite{curtarolo:art65}. 
Using a standard set of calculation parameters~\cite{curtarolo:art104},
input files are automatically generated for the
\VASP~\cite{kresse_vasp, vasp} DFT software package
with projector-augmented-wave pseudopotentials~\cite{PAW} and the PBE parameterization of the
generalized gradient approximation to the exchange-correlation functional~\cite{PBE}. 
Calculations are monitored as they run to detect and correct for errors without the 
need for any user intervention.
Useful materials data is then extracted and processed for 
dissemination through the \AFLOW.org repository.
The entire framework is written in the \texttt{C++} programming language (more than highly integrated 400,000 lines, as of version 3.1.153), providing
a robust platform for continuous infrastructure development with
reliable high performance.
\subsection{AFLOW: Automated Workflows}

\AFLOW\ offers several automated workflows, each dedicated to a specific type of 
characterization yielding a set materials properties.
For electronic properties, \AFLOW\ performs four DFT calculations:
two rounds of geometry relaxation (stage name: ``RELAX'') using
the \VASP\ conjugate gradient optimization algorithm,
a static run (\textit{i.e.}, fixed geometry; stage name: ``STATIC'') with a denser \textbf{k}-point mesh to obtain
an accurate density of states,
and a band structure calculation (stage name: ``BANDS'') following the \AFLOW\ Standard path through the
high-symmetry \textbf{k}-points in the Brillouin Zone~\cite{curtarolo:art58}.

Other workflows in \AFLOW\ manage ensembles of DFT calculations,
all offering the same automated error-correction procedures for high-throughput processing.
For thermal and elastic properties, the Debye-Gr{\"u}neisen model 
(\underline{A}utomatic \underline{G}IBBS \underline{L}ibrary, \AGL)~\cite{curtarolo:art96}
is combined with the \underline{A}utomatic \underline{E}lasticity \underline{L}ibrary (\AEL)~\cite{curtarolo:art115}
as described in Section~\ref{sec:aelagl}.
A more accurate thermal characterization can be resolved with the finite displacement method for phonon calculations
(\underline{A}utomatic \underline{P}honon \underline{L}ibrary, \APL)~\cite{curtarolo:art114} and
its associated extensions, \textit{i.e.}, the \underline{q}uasi-\underline{h}armonic \underline{a}pproximation (\QHAAPL)~\cite{curtarolo:art114}
and \underline{A}utomatic \underline{A}nharmonic \underline{P}honon \underline{L}ibrary (\AAPL)~\cite{curtarolo:art125},
as described in Section~\ref{sec:apl}.
\AFLOW\ also extends beyond ideal crystalline materials characterization, offering
modules to investigate off-stoichiometric materials (\AFLOWPOCC, Section~\ref{sec:pocc})~\cite{curtarolo:art110}
and to predict metallic glass formation as a function of composition~\cite{curtarolo:art112}.
\subsection{AFLOW: Prototype Library}

The \AFLOW\ framework uses decorated crystal structure prototypes for materials discovery \cite{curtarolo:art121}.
Structural prototypes are specific arrangements of atoms which are commonly observed in nature, such as the rocksalt,
zincblende and wurtzite structures.
The atomic sites in these prototypes are populated with different elemental species to generate 
materials structures, for which the properties and thermodynamic stability are then obtained from DFT calculations.
An extensive list of the structural prototypes included in \AFLOW\ has been published in Ref. \onlinecite{curtarolo:art121}
and is available online at \verb|http://aflow.org/CrystalDatabase|.

Pages within the website display a curated list of data for each structural prototype, including materials exhibiting
this structure, various symmetry descriptions, the primitive and atomic basis vectors, and
original references where the structure was observed.
Accompanying these descriptions is an interactive \verb|Jmol| visualization of the prototype, as described in 
Section~\ref{sec:jmol}.
The page also contains a prototype generator, where the structural degree(s) of freedom and atomic species are defined to create
new materials by leveraging the \AFLOW\ prototypes module.
This generates the corresponding input file for one of many {\it ab initio} software packages, including
\VASP~\cite{kresse_vasp, vasp}, \QUANTUMESPRESSO~\cite{qe,
  Giannozzi:2017io}, \ABINIT~\cite{gonze:abinit}, and
\FHIAIMS~\cite{blum:fhi-aims}.

\subsection{AFLOW-SYM: Symmetry Analyzer}
\label{sec:aflowsym}

\begin{figure}
  \begin{center}
    \includegraphics[width=0.5\textwidth]{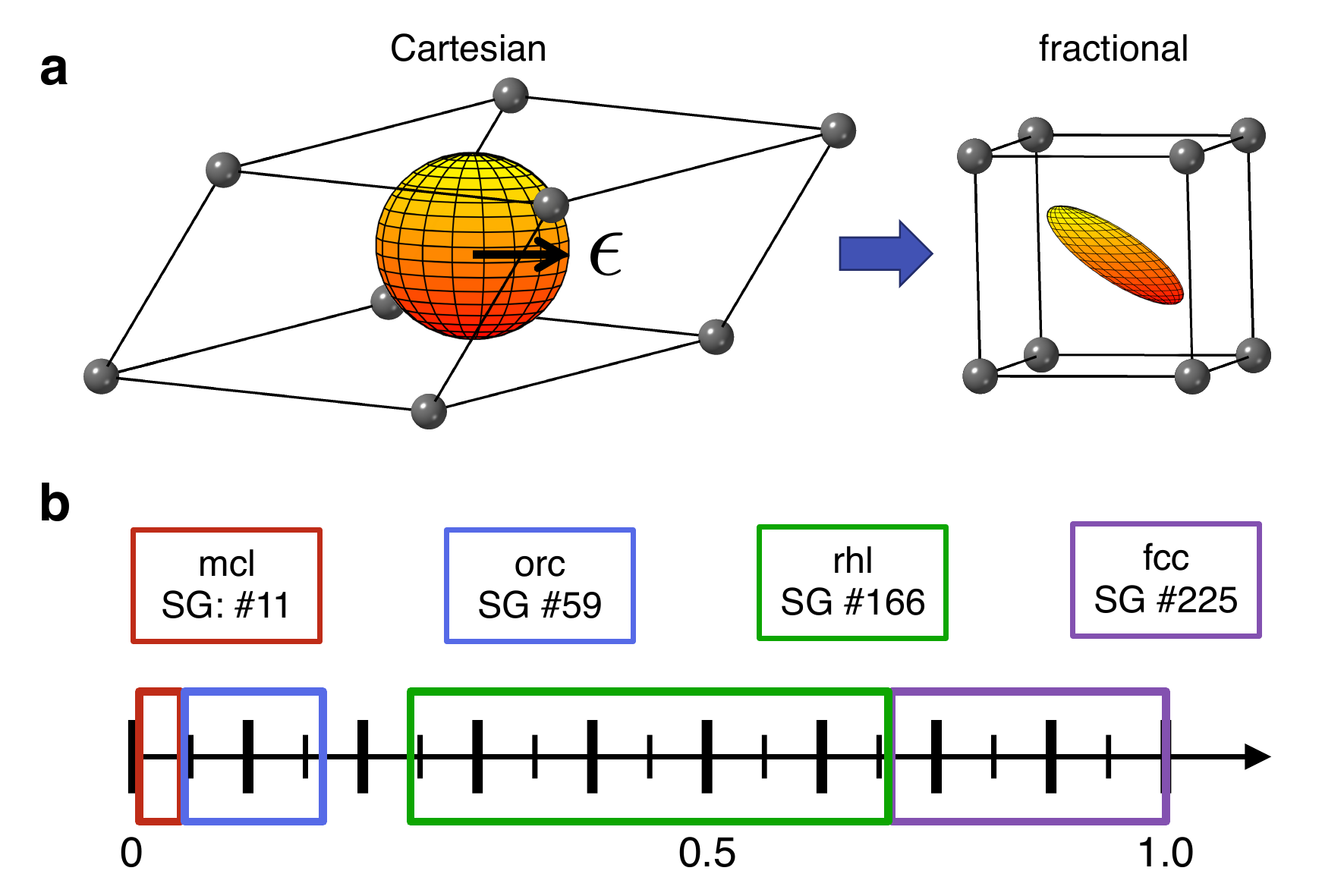}
    \caption{\textbf{Visualization of tolerance-sphere warping and adaptive tolerance method.} 
    (\textbf{a}) Illustration of the warping of space when transforming from cartesian to fractional coordinates in the general case.
    (\textbf{b}) Spectrum of space groups identified by \AFLOWSYM\ with different tolerance choices.
    }
    \label{fig:tol_and_scan}
  \end{center}
\end{figure}

The \AFLOW\ framework automatically analyzes the symmetry of materials structures, and returns a complete symmetry description.
To address numerical tolerance issues, \AFLOW\ employs an atom mapping
procedure that is reliable even for non-orthogonal unit cells, and
uses an adaptive tolerance scheme to ensure symmetry results are commensurate with crystallographic principles (see 
Figure~\ref{fig:tol_and_scan}).
These routines --- referred to as \AFLOWSYM~\cite{aflowsym_2017} --- are robust, and have been used to successfully determine the symmetry 
properties of over 1.7 million materials in the \AFLOW\ repository.

Structural isometries are identified by determining the set of symmetry operators that lead to 
isomorphic mappings between the original and transformed atoms. 
The structure exhibits symmetry under a particular operation if the set of closest mapping 
distances are all below a tolerance threshold $\epsilon_{0}$. 
Periodic boundary conditions introduce complexity for finding the minimum mapping vector, 
necessitating the exploration of neighboring cells.  
This is achieved via the method of images through either 
(\textit{i}) a unit cell expansion, yielding the globally optimal distance or 
(\textit{ii}) a bring-in-cell method (generally performed in fractional coordinates) that 
reduces each component of the distance vector independently. 
While computationally inexpensive compared to the unit cell expansion, the bring-in-cell method 
is only exact for orthogonal lattices (i.e. described by a diagonal metric tensor), since it 
does not consider overlap between lattice vectors (see Figure~\ref{fig:tol_and_scan}(a)). 
To safely exploit the bring-in-cell approach, \AFLOWSYM\ employs a heuristic maximum tolerance 
$\epsilon_{\mathrm{max}}$ based on the maximum lattice skewness with a threshold 
which guarantees consistent and accurate results~\cite{aflowsym_2017}.

Given a particular tolerance value, different symmetry operations can be realized in or excluded 
from the description of a crystal.
Figure~\ref{fig:tol_and_scan}(b) highlights how the tolerance value affects the possible space groups
for AgBr (\ICSD\ \#56551 with a reported space group \#11).
The neighboring space group regions are consistent with non-isomorphic subgroup relations, namely 
between space groups \#59 and \#11 and between \#225 and \#166.
However, a gap or ``confusion'' tolerance region occurs between space groups \#59 and \#166 (with no direct 
subgroup relations).
The problematic regions stem from noise in the structural data, impeding the identification of operations consistent with symmetry principles.
This problem is solved by using a radial tolerance scan extending from the input tolerance
$\epsilon_{0}$.
Given a change in tolerance, the algorithm recalculates and verifies all symmetry properties
until a globally consistent description is identified.

\AFLOWSYM\ is compatible with many established {\it ab initio} input files, including those for \VASP~\cite{vasp},
\QUANTUMESPRESSO~\cite{qe, Giannozzi:2017io}, \ABINIT~\cite{gonze:abinit}, and \FHIAIMS~\cite{blum:fhi-aims}.
From the structural information, \AFLOWSYM\ delivers the symmetries of the lattice, crystal (lattice + atoms),
reciprocal lattice, superlattice (equally decorated sites), and crystal-spin (lattice + atoms + magnetic moment).
This affords a multitude of symmetry descriptions to be presented, such as the space group number/symbol(s), 
Pearson symbol, point group symbol(s), Wyckoff positions, and Bravais lattice type/variation~\cite{curtarolo:art58}.
Moreover, the operators of the different symmetry groups --- including the point groups, factor groups, 
space group, and site symmetries --- are provided to users in rotation matrix, axis-angle, matrix generator, and quaternion 
representations for easy manipulation.  
All symmetry functions support the option to output in \JSON\ format.
This allows \AFLOWSYM\ to be leveraged from other programming languages such as Java, Go, Ruby, Julia and Python --- facilitating the
incorporation of \AFLOWSYM\ into numerous applications and workflows.

\subsection{\AFLOW-HULL: Convex Hull Analysis}
\label{sec:aflowchull}

\begin{figure*}
  \includegraphics[width=\textwidth]{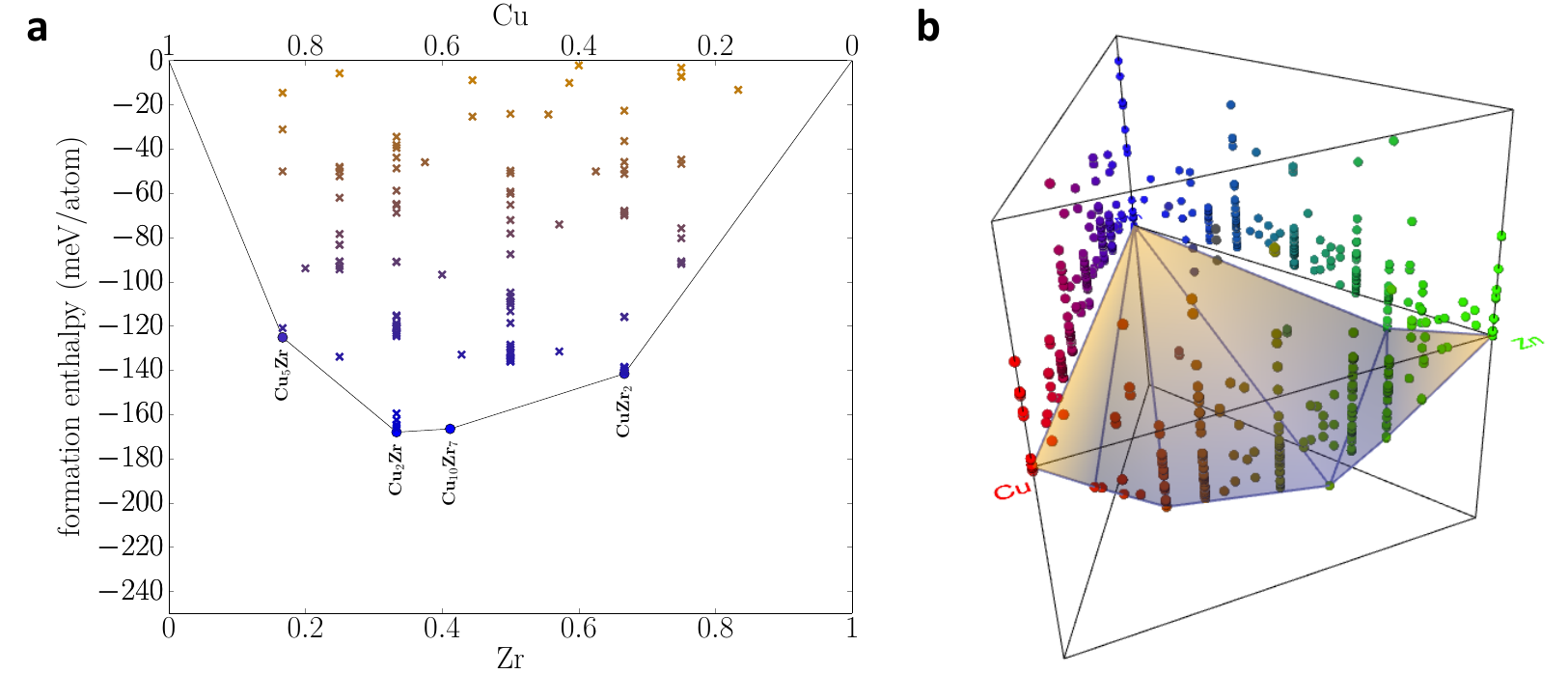}
  \caption{\small\textbf{Example convex hull illustrations offered by \AFLOW.}
    (\textbf{a}) 2D convex hull of the CuZr system generated automatically by \AFLOW.
    (\textbf{b}) 3D convex hull of the CuMnZn system presented through the \AFLOWHULL\ application
    online: \url{http://aflow.org/aflow-hull}.
  }
  \label{fig:hull}
\end{figure*}

Structure and energy data from the \AFLOW.org repository are used
to resolve the low-temperature/low-pressure thermodynamic stability of
compound systems.
For a given stoichiometry, the \AFLOW.org repository provides the DFT energies of various crystal polymorphs.
By exploring representative structures over the full range of stoichiometries,
\AFLOWHULL\ constructs the minimum energy surface, \textit{i.e.}, the lower-half convex hull~\cite{qhull},
defining thermodynamic stability for the system (at zero temperature and pressure).
Structures on the hull are thermodynamically stable (ground state), while those far from the hull
will decompose into a combination of stable phases, dictating synthesizability at these conditions.
Any analysis of the hull requires sufficient statistics to ensure convergence,
\textit{i.e.}, enough representative structures have been included in
the alloy system calculations such that any additional entries are not expected to change the minimum energy surface.

The geometric construction offers several key properties critical for synthesizability.
For a specific composition, the energetic distance to the hull quantifies the energy released during the decomposition,
while the ground state phases defining the tie-line/facet below the compound 
are the products of the reaction.
The distance from the hull also measures the ``severity'' of instability, \textit{i.e.},
structures near the hull may stabilize at higher temperatures or pressures.
Similarly, a robust stability criterion 
can be quantified for ground state phases by removing
the phase from the set and measuring the distance of the compound from the new hull.
The larger the distance, the less likely the ground state phase will become unstable
at higher temperatures/pressures~\cite{curtarolo:art109}.
The generalized tie-lines (facet ridges) dictate which phases can coexist in equilibrium,
and play a role in determining the feasibility of synthesis/treatment techniques, such as
precipitation hardening~\cite{curtarolo:art113}.

Given a compound system, \AFLOWHULL\ automatically queries the \AFLOW.org database, constructs
the hull, calculates the aforementioned properties, and delivers the information
in one of the following formats: PDF, plain-text, and \JSON.
\AFLOWHULL\ can also visualize the 2D and 3D hulls, as illustrated in Figure~\ref{fig:hull}.
In the case of the PDF output format, hyperlinks are included to allow 
for additional queries of the full properties set offered through the \AFLOW.org repository.
Links are also added connecting the hull visualization to relevant properties
for easy navigation of the full PDF document.

A separate online application, available at \verb|aflow.org/aflow-hull|, has been created to showcase the results of \AFLOWHULL,
and provides interactive binary and ternary convex hull visualizations. 
The application consists of four components: the periodic table, visualization viewport, selected entries list, and the comparison page. 
The periodic table component is the entry point of the application and provides the interface to search for convex hulls of different alloy systems. 
Elements within the periodic table respond when selected to display information to the user. 
As a selection is made, the color of each border will change to green, yellow, and red based on hull reliability.
A reliability threshold of 200 compounds for a binary hull has been heuristically defined.
Selections highlighted in green are well above this threshold, while those in yellow/red are near/below the cutoff.

When a hull is selected, the application transitions to the visualization viewport component. 
Depending on the number of elements selected, a 2D plot (binary) or 3D plot (ternary) will appear. 
Each plot is interactive, allowing points to be selected, where each point represents an entry in the \AFLOW\ repository. 
Information for each point is displayed in the selected entries list component, 
which is accessible through the navigation bar.
Selected hulls will appear on the comparison page as a grid of cards, and selected points are highlighted across 
all hulls containing those entries.

\subsection{AFLOW-POCC: Partial Occupations}
\label{sec:pocc}

\begin{figure*}
  \includegraphics[width=\textwidth]{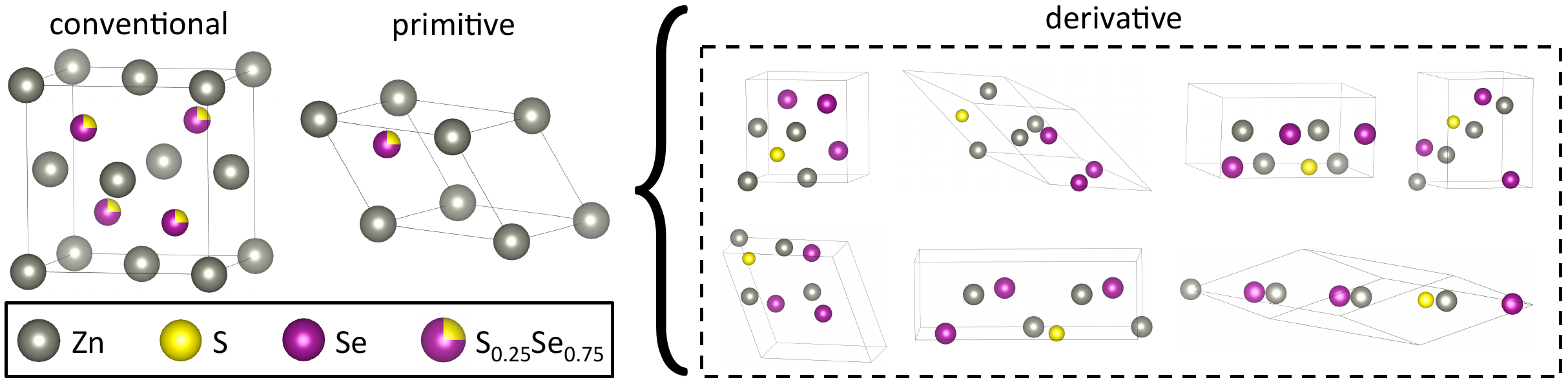}
  \caption{\small\textbf{Structure enumeration for off-stoichiometric materials modeling.}
    For the off-stoichiometric material ZnS$_{0.25}$Se$_{0.75}$, a superlattice of size $n=4$ accommodates the stoichiometry exactly.
    By considering all possibilities of decorated supercells and eliminating duplicates by UFF energies, seven structures are identified as unique.
    These representative structures are fully characterized by \AFLOW\ and \VASP, and are ensemble-averaged to resolve the system-wide properties.
  }
  \label{fig:pocc}
\end{figure*}

The \AFLOW\ \underline{P}artial \underline{Occ}upation module (\AFLOWPOCC) \cite{curtarolo:art110} models configurational 
and structural disorder including substitutions, vacancies,
and random lattice site occupation, by generating a set of representative configurations. 
First, a supercell size is determined that accommodates the fractional stoichiometry to within a user-defined tolerance.
Given the supercell size, $n$, superlattices are generated using Hermite Normal Form
matrices~\cite{gus_enum1}, which are then decorated 
in accordance with the stoichiometry to generate all possible configurations,
as illustrated in Figure~\ref{fig:pocc}.
Duplicate configurations are rapidly identified and eliminated by estimating
the energy of each structure using the \underline{U}niversal \underline{F}orce \underline{F}ield (UFF) model~\cite{Rappe_1992_JCAS_UFF}.
The properties of the remaining unique configurations are calculated with DFT, and 
ensemble-averaged to resolve system-wide properties of the disordered material.
The ensemble-average employs a Boltzmann distribution weight which is a function of a disorder parameter (temperature), 
energy relative to the ground state configuration, and degeneracy as determined by the UFF model.
Ensemble-average properties include the electronic band gap, density of states, and magnetic moment.
\subsection{AEL and AGL: Thermo-mechanical Properties}
\label{sec:aelagl}

The \AFLOW\ \underline{A}utomatic \underline{E}lasticity \underline{L}ibrary (\AFLOW-\AEL\ \cite{curtarolo:art115}) and the 
\AFLOW\ \underline{A}utomatic \underline{G}IBBS \underline{L}ibrary (\AFLOW-\AGL\ \cite{curtarolo:art96}) modules determine 
thermo-mechanical materials properties from calculations of strained primitive cells. 
These methods are generally computationally less costly than the phonon (\APL\ and \AAPL) calculations described in Section \ref{sec:apl}, 
although \APL\ and \AAPL\ generally give more quantitatively accurate results, particularly for properties where anharmonic effects are important. 
\AEL\ and \AGL\ have been combined into a single automated workflow, which has been 
used to calculate the thermo-mechanical properties for over 5000 materials in the \AFLOW\ repository.

\begin{figure*}
  \includegraphics[width=0.98\textwidth]{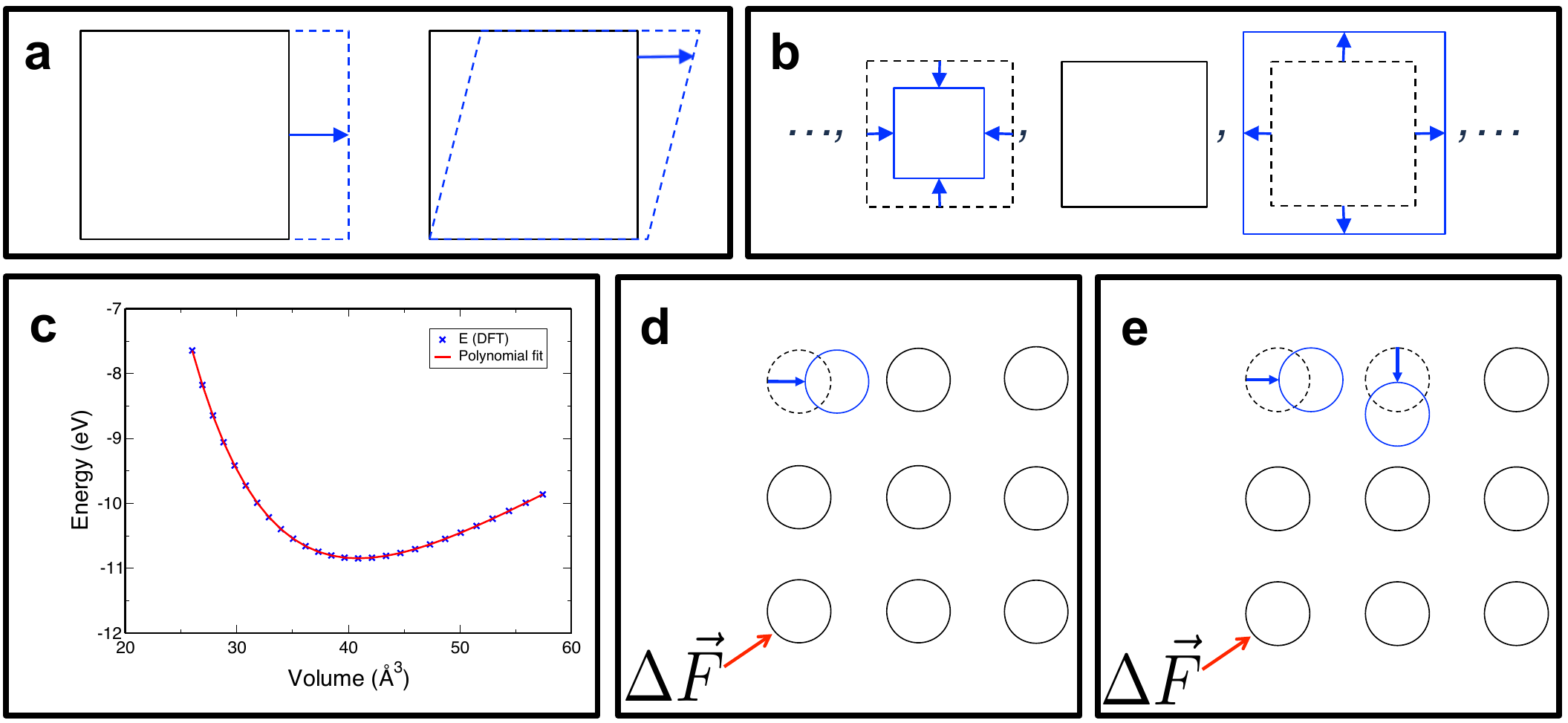}
  \vspace{-4mm}
  \caption{\small {\bf Ensembles of distorted geometries for the calculation of thermo-mechanical properties with \AFLOW.}
    \AEL\ uses the stresses from a set of (\textbf{a}) normal and shear strained cells to obtain the elastic constants.
   \AGL\ calculates the energies of a set of  (\textbf{b}) isotropically compressed and expanded unit cells, and  (\textbf{c}) fits the 
   resulting $E(V)$ data by a numerical polynomial or by an empirical equation of state.    
   \APL\ obtains the (\textbf{d}) second order harmonic IFCs from a set of single atom displacements and the 
   (\textbf{e}) third order anharmonic IFCs from a set of 2-atom displacements.}
  \label{fig:distorted_geometry}
\end{figure*}

The \AEL\ module  applies a set of independent normal and shear strains to the primitive cell of a material \cite{curtarolo:art100, curtarolo:art115} 
as depicted in Figure \ref{fig:distorted_geometry}(a), and uses DFT to calculate the resulting stress tensors. 
This set of strain-stress data is used to generate the elastic stiffness tensor, {\it i.e.} the elastic constants:
\begin{equation}
\left( \begin{array}{l} \sigma_{11} \\ \sigma_{22} \\ \sigma_{33} \\ \sigma_{23} \\ \sigma_{13} \\ \sigma_{12} \end{array} \right) = 
\left( \begin{array}{l l l l l l} c_{11}\ c_{12}\ c_{13}\ c_{14}\ c_{15}\ c_{16} \\ 
 c_{12}\ c_{22}\ c_{23}\ c_{24}\ c_{25}\ c_{26} \\ 
c_{13}\ c_{23}\ c_{33}\ c_{34}\ c_{35}\ c_{36} \\ 
c_{14}\ c_{24}\ c_{34}\ c_{44}\ c_{45}\ c_{46} \\ 
c_{15}\ c_{25}\ c_{35}\ c_{45}\ c_{55}\ c_{56} \\ 
c_{16}\ c_{26}\ c_{36}\ c_{46}\ c_{56}\ c_{66} \end{array} \right)
\left( \begin{array}{c} \epsilon_{11} \\ \epsilon_{22} \\ \epsilon_{33} \\ 2\epsilon_{23} \\ 2\epsilon_{13} \\ 2\epsilon_{12} \end{array} \right)
\end{equation}
written in the $6 \times 6$ Voigt notation using the mapping \cite{Poirier_Earth_Interior_2000}: 
$11 \mapsto 1$, $22 \mapsto 2$, $33 \mapsto 3$, $23 \mapsto 4$, $13 \mapsto 5$, $12 \mapsto 6$. These elastic constants are
combined to calculate the bulk, $B$, and shear, $G$, elastic moduli in the Voigt, Reuss, and Voigt-Reuss-Hill (VRH, $B_\sVRH$ and $G_\sVRH$) 
approximations. The Poisson ratio $\nu$ is then given by:
\begin{equation}
  \label{Poissonratio}
  \nu = \frac{3 B_{\sVRH} - 2 G_{\sVRH}}{6 B_{\sVRH} + 2 G_{\sVRH}}.
\end{equation}

The \AGL\ module is based on the \GIBBS\ \cite{Blanco_CPC_GIBBS_2004, Blanco_jmolstrthch_1996}  quasi-harmonic Debye-Gr{\"u}neisen 
model, and calculates the energy as a function of volume, $E(V)$, for a set of isotropically compressed and expanded strains of the 
primitive cell, as illustrated in Figure \ref{fig:distorted_geometry}(b). The $E(V)$ data are fitted by either a numerical polynomial 
or an empirical equation of state to obtain the adiabatic bulk modulus $B_\sS(V)$, as shown in Figure \ref{fig:distorted_geometry}(c). 
The Debye temperature $\theta_\sD(V)$ as a function of volume is then calculated using the expression:
\begin{equation}
  \label{debyetemp}
  \theta_\sD = \frac{\hbar}{k_\sB}[6 \pi^2 V^{1/2} n]^{1/3} f(\nu) \sqrt{\frac{B_\sS}{M}},
\end{equation}
where $n$ is the number of atoms per unit cell, $M$ is the unit cell mass, and $f(\nu)$ is a function of the Poisson ratio $\nu$:
\begin{equation}
  \label{fpoisson}
  f(\nu) = \left\{ 3 \left[ 2 \left( \frac{2}{3} \!\cdot\! \frac{1 + \nu}{1 - 2 \nu} \right)^{3/2} \!\!\!\!\!\!\!+ \left( \frac{1}{3} \!\cdot\! \frac{1 + \nu}{1 - \nu} \right)^{3/2} \right]^{-1} \right\}^{\frac{1}{3}}\!\!\!\!,
\end{equation}
where $\nu$ can be obtained from Equation \ref{Poissonratio} using \AEL, or set directly by the user. The vibrational contribution to the free energy, $F_\svib$, is given by:
\begin{equation}
  \label{helmholtzdebye}
  F_\svib(\theta_\sDebye; T) \!=\! n k_\sB T \!\left[ \frac{9}{8} \frac{\theta_\sDebye}{T} \!+\! 3\ \mathrm{log}\!\left(1 \!-\! {\mathrm e}^{- \theta_\sDebye / T}\!\right) \!\!-\!\! D\left(\frac{\theta_\sDebye}{T}\right)\!\!\right],
\end{equation}
where $D(\theta_\sDebye / T)$ is the Debye integral:
\begin{equation}
  D \left(\theta_\sDebye/T \right) = 3 \left( \frac{T}{\theta_\sDebye} \right)^3 \int_0^{\theta_\sDebye/T} \frac{x^3}{e^x - 1} dx.
\end{equation}
The Gibbs free energy is obtained from:
\begin{equation}
  \label{gibbsdebye}
  {\sf G}(V; p, T) = E_\sDFT(V) + F_\svib (\theta_\sD(V); T)  + pV.
\end{equation}
The volume which minimizes $ {\sf G}(V; p, T)$ at a given pressure $p$ and temperature $T$ is the equilibrium volume $V_{\mathrm{eq}}$, which 
is used to evaluate $\theta_\sD (V_{\mathrm{eq}})$ and the Gr{\"u}neisen parameter $\gamma$ as defined by:
\begin{equation}
  \label{gruneisen_theta}
  \gamma = - \frac{\partial \ \mathrm{log} (\theta_\sD(V))}{\partial \ \mathrm{log} V}.
\end{equation}
Finally, $\theta_\sD$ and $\gamma$ are used to calculate other thermal properties including $C_\sV$, $C_{\mathrm p}$, 
$\alpha_\sV$ and $\kappa_{\mathrm L}$ \cite{curtarolo:art96, Blanco_CPC_GIBBS_2004}.

\subsection{AFLOW-APL: Phonons}
\label{sec:apl}

The \AFLOW\ \underline{A}utomatic \underline{P}honon \underline{L}ibrary (\AFLOW-\APL)~\cite{curtarolo:art65} 
calculates the harmonic vibrational properties of a crystal using the finite displacement method.
Computed properties include the phonon dispersion and density of states, 
vibrational entropy $\left(S_{\mathrm{vib}}\right)$,
and the heat capacity (at constant volume, $C_{\mathrm{V}}$) as a function of temperature.
These features are determined through an analysis of the phonon modes, accessed through the 
\underline{I}nteratomic \underline{F}orce \underline{C}onstants (IFCs)~\cite{Maradudin1971}.
To first approximation, the harmonic (second order) IFC $C_{\substack{i,j;\alpha,\beta}}$ 
is the negative of the force exerted
in the $\alpha$ direction on the atom $i$ when the atom
$j$ is displaced in the $\beta$ direction, with all other atoms maintaining their equilibrium position.
To determine the forces, the atoms of the structure (supercell) are individually 
 perturbed as illustrated in Figure~\ref{fig:distorted_geometry}(d).
The forces are obtained with DFT from the derivative of the total energy using the Hellmann-Feynman theorem.
Supercells are used to sufficiently capture/isolate the impact of the distortion on the structure;
distortions on small cells create forces on all atoms as well as their periodic images.

Given an input structure, \AFLOW\ creates the full set of distorted supercell structures
for the calculation of the forces.
To minimize the number of expensive DFT calculations (primary
computational bottle-neck), \AFLOWSYM\ (see Section \ref{sec:aflowsym}) is employed
to determine which distortions are symmetrically equivalent using the site symmetry.
Only inequivalent distortions are applied and explicitly calculated.
Symmetry is then used to appropriately construct the IFC matrix, from which
the dynamical matrix is constructed and the phonon modes, energies, and group velocities are derived.

\APL\ has been extended to include the calculation of quasi-harmonic (\underline{q}uasi-\underline{h}armonic \underline{a}pproximation \APL, \QHAAPL\ 
\cite{curtarolo:art114, curtarolo:art119}) and anharmonic (\underline{A}utomatic \underline{A}nharmonic \underline{P}honon \underline{L}ibrary, 
\AAPL\ \cite{curtarolo:art125}) effects in order to obtain properties such as the heat capacity at constant pressure $C_{\mathrm{p}}$, 
coefficient of volumetric thermal expansion $\alpha_{\mathrm V}$, and lattice thermal conductivity $\kappa_{\mathrm L}$. 

\QHAAPL\ performs harmonic \APL\ calculations at multiple different volumes, and extracts the Gr{\"u}neisen parameter from the change of the 
phonon frequencies with respect to volume:
\begin{equation}
\label{gamma_micro}
 \gamma_i = - \frac{V}{\omega_i} \frac{\partial \omega_i}{\partial V}.
\end{equation}
The Gr{\"u}neisen parameter can be used in combination with harmonic properties such as $C_{\mathrm V}$ to calculate $C_{\mathrm p}$, 
$\alpha_{\mathrm V}$ \cite{curtarolo:art114} and $\kappa_{\mathrm L}$ \cite{curtarolo:art119}.

\AAPL\ obtains the third order anharmonic IFCs  by distorting two atoms in a supercell structure at a time as depicted in 
Figure \ref{fig:distorted_geometry}(e), and then calculating the change in forces on the other atoms \cite{curtarolo:art125}. 
These IFCs are used to calculate the three-phonon scattering rates, and thus the scattering time and mean free displacement. 
These quantities are combined with the group velocities obtained from harmonic \APL\ to solve the Boltzmann Transport Equation and 
calculate $\kappa_{\mathrm L}$ with quantitative accuracy \cite{curtarolo:art125}.
\subsection{AFLOW: Visualization Tools}
\label{sec:jmol}

\AFLOW\ leverages a panoply of visualization tools for materials data,
including standard software such as \verb|gnuplot|, \verb|latex|, and \verb|xmgrace|
for plots of phonon dispersions, electronic band structures, electronic density of states, and convex hull visualization.
These plots are served publicly through the \AFLOW.org repository.

\begin{figure}
  \includegraphics[width=\linewidth]{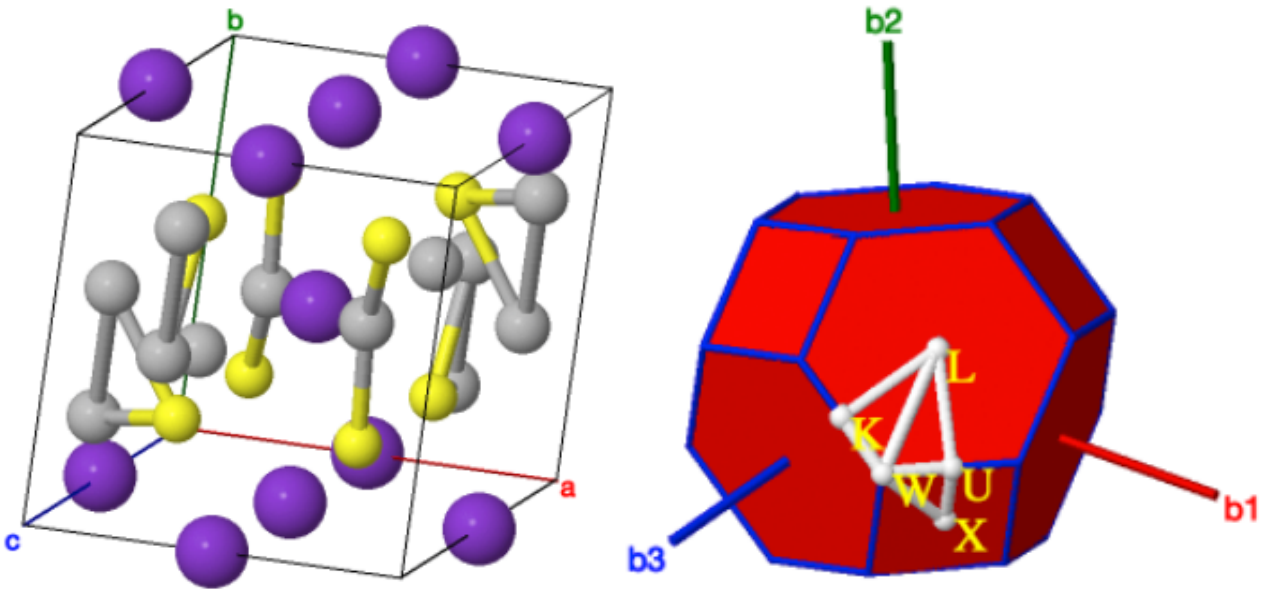}
  \caption{\small\textbf{Side-by-size visualization of the crystal structure and Brillouin Zone using Jmol.}
    The structure highlighted is Ag$_{3}$KS$_{2}$ (ICSD \#73581): \url{http://aflow.org/material.php?id=Ag6K2S4_ICSD_73581}.
    The \AFLOW\ Standard path of high-symmetry \textbf{k}-points is illustrated in the Brillouin Zone~\cite{curtarolo:art58}.
  }
  \label{fig:bzplotter}
\end{figure}

To visualize crystal structures, \AFLOW\ employs the \verb|Jmol| software, which
has incorporated substantial functionality for \AFLOW-specific application.
The \verb|JSmol| branch of the software powers the online crystal structure visualizations in
the \AFLOW.org repository entry pages and \AFLOW\ Prototype Library pages.
With its recently added POSCAR reader, \verb|JSmol| provides an assortment of capabilities ranging from
different view perspectives, supercell expansions, and varying unit cell representations.
A similar visualization application showing the \AFLOW\ Standard high-symmetry paths
in the Brillouin Zone \cite{curtarolo:art58} is currently being incorporated, as illustrated in Figure~\ref{fig:bzplotter}.
Additionally, the \verb|Jmol| desktop client offers a specialized macro (\texttt{aflow}) for visualization
of alloy systems, which leverages the \AFLUX\ Search-{\small API}.

\section{AFLOW$\pi$: Minimalist high-throughput}
\label{sec:aflowpi}

\begin{figure*}
  \includegraphics[width=0.98\textwidth]{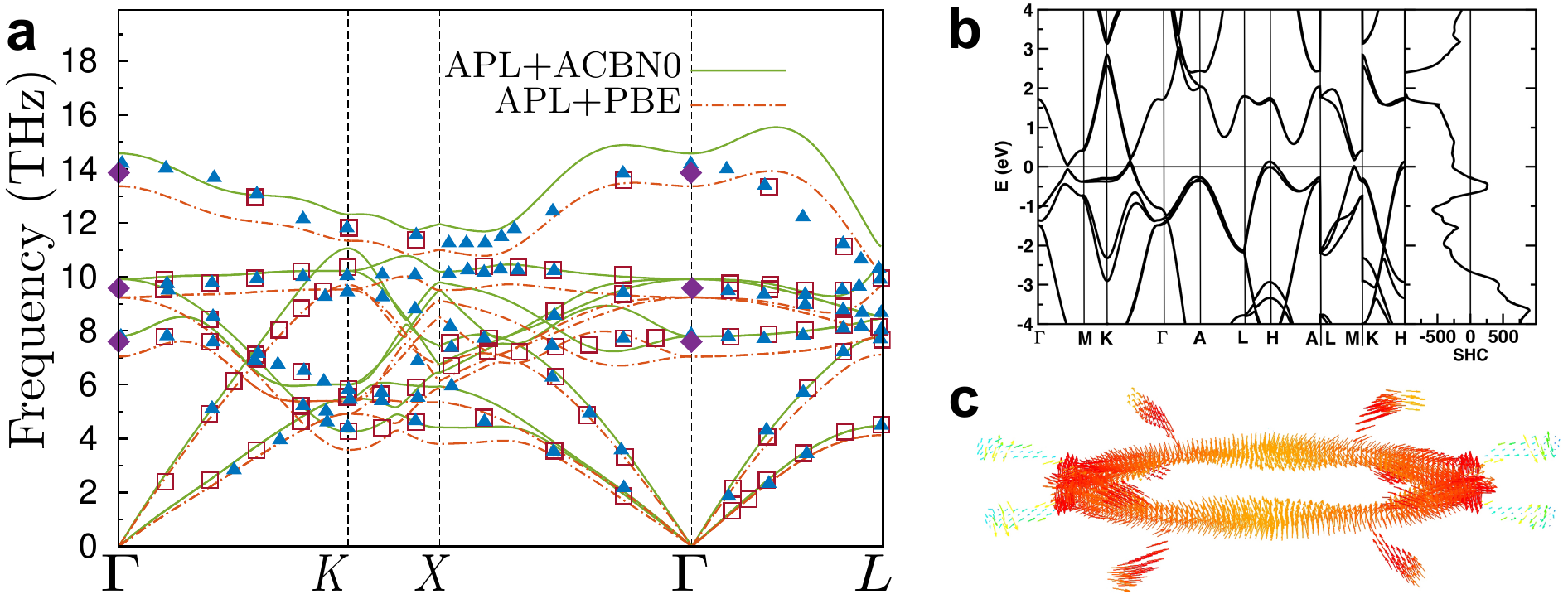}
  \vspace{-4mm}
  \caption{\small {\bf Vibrational spectrum calculated with \AFLOWpi\ and \AFLOW-\APL\ (left) and electronic properties
  computed with \PAOFLOW\ (right).}
     (\textbf{a}) Phonon dispersion of CaF$_2$ calculated with \APL\ \cite{curtarolo:art125}, using the ACBN0 method as implemented within \AFLOWpi\
     (green lines). The results obtained using PBE are shown by the broken orange lines for comparison.
     The blue triangles and red unfilled squares represent neutron scattering data from Ref. \onlinecite{Schmalzl_prb_2003}
     and Ref. \onlinecite{Elcombe_jpcss_1970} respectively, while the purple diamonds represent Raman and infrared data from Ref. \onlinecite{Kaiser_pr_1962}.
     (\textbf{b}) Electronic band structure, \underline{s}pin \underline{H}all \underline{c}onductivity (SHC) and
     (\textbf{c}) spin texture of the nodal line and Weyl points in HfC, as calculated using the PAOFLOW utility.}
  \label{fig:aflowpi_phonons_paoflow}
\end{figure*}

The \AFLOWpi\ \cite{curtarolo:art127} framework has been originally implemented as a minimalist software to perform
verification tasks (see Section \ref{sec:valver}) on data published on \AFLOW.org. By design, \AFLOWpi\ is easy to
install and to extend to a variety of
electronic structure codes (currently only the \QUANTUMESPRESSO\ \cite{qe, Giannozzi:2017io} DFT package is implemented).
\AFLOWpi\ builds on the versatility of Python, providing a module to prepare, run, and analyze large sets of first
principles calculations,
and includes tools for the automatic \underline{p}rojection on pseudo-\underline{a}tomic \underline{o}rbitals
(PAO, see Section \ref{sec:paoflow}),
and the self-consistent calculation of Hubbard $U$ corrections within the \underline{A}gapito, \underline{C}urtarolo
and \underline{B}uongiorno
\underline{N}ardelli (ACBN0) approach \cite{curtarolo:art93, Andrade_Octopus_PCCP_2015}. In addition, workflows for
the calculation of elastic constants, diffusive transport coefficients, optical spectra and
phonon dispersions
with DFT+U (see Figure \ref{fig:aflowpi_phonons_paoflow}(a) for assessing the effect of the Hubbard $U$ corrections on
the phonon dispersion
calculated using \APL). When possible, \AFLOWpi\ exploits the tight-binding hamiltonians as in Ref. \onlinecite{curtarolo:art116}.
Calculation results can be easily packaged and prepared for incorporation into the \AFLOW.org data repository (see Section \ref{sec:reposit}).

\section{PAOFLOW: Fast characterization}
\label{sec:paoflow}

\PAOFLOW\ \cite{paoflow} is a stand-alone tool to efficiently post-process standard DFT pseudo-potential plane-wave calculations to generate
\underline{t}ight-\underline{b}inding (TB) hamiltonians which faithfully reproduce the calculated electronic structure (eigenvalues and eigenvectors)
with arbitrary precision in reciprocal space \cite{curtarolo:art86,curtarolo:art111,curtarolo:art108} (see Figure \ref{fig:aflowpi_phonons_paoflow}(b) for
\PAOFLOW\ generated band structure for HfC  (\ICSD\ \#169399, space group \#187, \AFLOW\ prototype: \verb|AB_hP2_187_d_a|)). 
By exploiting the simplicity of the TB formalism and the efficiency of fast Fourier transforms, \PAOFLOW\ interpolates
the band structure and computes the matrix elements of the momentum operator, ${\bf p}_{m,n}$. 
These are used to improve the quality of integrated quantities such as the density of states (adaptive smearing), to compute electronic transport coefficients 
within the constant relaxation time approximation, and to compute the dielectric constants \cite{curtarolo:art116}. 
In addition, the ${\bf p}_{m,n}$ matrix elements facilitate the calculation of the Berry curvature and related properties 
(anomalous Hall conductivity, spin Hall conductivity (see Figure \ref{fig:aflowpi_phonons_paoflow}(b)), 
magnetic circular dichroism, spin circular dichroism; see spin texture of the nodal line and Weyl points in HfC shown in Figure \ref{fig:aflowpi_phonons_paoflow}(c)).
Starting from a well interpolated band structure, it also possible to compute topological invariants.

Because of the local representation of the electronic structure provided by the \PAOFLOW\ software, surface projected band structure and
Landauer ballistic transport are also computable within \PAOFLOW.

The software is implemented in Python, is portable and easy to install, and is parallel by design (on both CPUs and GPUs). 
\PAOFLOW\ is also an integral part of the \AFLOWpi\ framework.

\section{AFLOW: Data Repository}
\label{sec:reposit}

The \AFLOW\ data repository \cite{curtarolo:art75} contains the calculated properties for over 1.7 million materials entries, obtained using 
the \AFLOW\ framework. These properties are available through the \verb|aflow.org| web portal, which includes online search/sort
and data analysis applications. The repository is programmatically accessible through the \AFLOW\ Data {\small REST-API} \cite{curtarolo:art92} and 
the \AFLUX\ Search-{\small API} \cite{curtarolo:art128}.
 
\subsection{AFLOW: Web Portal}
\label{sec:webportal}

\begin{figure*}
  \includegraphics[width=0.98\textwidth]{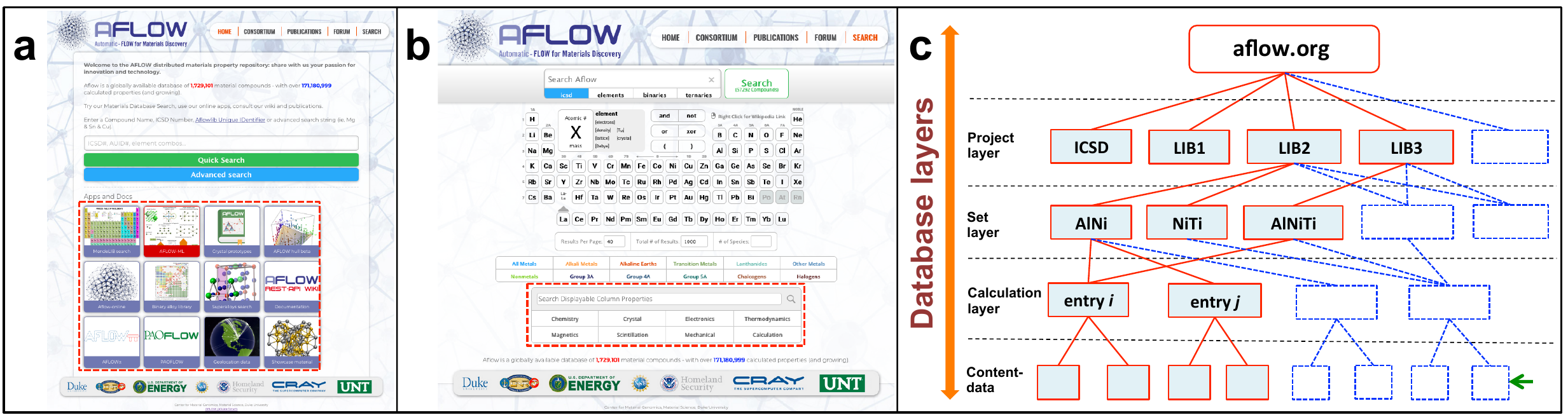}
  \vspace{-4mm}
  \caption{\small {\bf \AFLOW\ web portal and data repository.} ({\bf a}) Online applications and documentation are
    accessible via the ``Apps and Docs'' set of buttons surrounded by the dashed red rectangle.
    ({\bf b}) The advanced search application can be used to search for specific compositions, and also includes property search filters, as highlighted by the dashed red rectangle.
    ({\bf c}) The \AFLOW\ data repository is organized into project, set ({\it i.e.} alloy system) and calculation ({\it i.e.} materials entry) layers.
}
  \label{fig:aflow_webpage}
\end{figure*}

The \AFLOW\ repository \cite{curtarolo:art75} is available online via the \verb|aflow.org| web portal (Figure \ref{fig:aflow_webpage}(a)). 
It contains multiple online applications for data access, processing and visualization, 
including the advanced ``MendeLIB'' search application at \verb|http://aflow.org/advanced.php| 
which facilitates searching for materials entries with filters for elemental composition and calculated properties (see Figure \ref{fig:aflow_webpage}(b)), 
the interactive convex hull application at \verb|http://aflow.org/aflow-hull| (see Section \ref{sec:aflowchull}), 
the online machine learning model at \verb|http://aflow.org/aflow-ml| (see Section \ref{sec:aflowml}), 
and the \AFLOW\ online tool at \verb|http://aflow.org/aflow_online.html| which gives access to \AFLOW\ crystal structure analysis and processing functions. 
These applications and more are accessible from the main \verb|aflow.org| web page, via the ``Apps and Docs'' set of buttons highlighted by the dashed red rectangle in Figure \ref{fig:aflow_webpage}(a).

\subsection{AFLOW-ML: Online Machine Learning}
\label{sec:aflowml}

The \AFLOW\ \underline{m}achine \underline{l}earning (\AFLOWML) online application provides a user interface to leverage machine-learning models trained on \AFLOW\ data. 
It accepts a standard structure file (POSCAR or QE) and outputs predictions for properties such as the band gap, elastic moduli, heat capacity, Debye temperature, vibrational free energy, and thermal expansion coefficient. 
Additionally, structures within the AFLOW repository can be imported via the sidebar. 
This application provides an accessible medium to retrieve machine learning predictions without the need to install a software library or machine learning package. 

Currently, \AFLOWML\ supports two different machine-learning models. 
The first model, \underline{p}roperty-\underline{l}abeled \underline{m}aterials \underline{f}ragments \cite{curtarolo:art124}, \verb|plmf|, has been trained using data from the \AFLOW\ repository, and predicts properties such as the electronic band gap, specific heat capacities and bulk/shear moduli.
The second method is the \underline{m}olar \underline{f}raction \underline{d}escriptor model \cite{curtarolo:art129}, \verb|mfd|, which predicts vibrational properties such as vibrational free energy and entropy, and is based only on the chemical composition of the material.

The \AFLOWML\ {\small API} \cite{aflowmlapi} offers programmatic access to the \AFLOWML\ online application, and provides a simplified abstraction that facilitates leveraging powerful machine learning models. 
This distills the prediction process down to its essence: from a structure file, return a prediction. 
Using the {\small API} is a two step process: first a structure file, in POSCAR 5 format (structure input for version 5 of \VASP), is posted ({\it i.e.} uploaded) to the endpoint \verb|/<model>/prediction| on the \verb|aflow.org| server
using standard HTTP libraries or dedicated programs such \verb|curl| or \verb|wget|, where \verb|<model>| specifies the machine learning model to use in the prediction (current options: \verb|plmf| and \verb|mfd|). 
When a prediction is submitted, a \JSON\ response object is returned that includes a task id. 
The results of the prediction are then retrieved from the \verb|/prediction/result/| endpoint on the \verb|aflow.org| server by appending the task id to the end of the URL, \textit{i.e.} \verb|/prediction/result/{id}/|. 
This endpoint monitors the prediction task and responds with a \JSON\ object that details its status. 
When complete, the endpoint responds with the results of the prediction, represented as a \JSON\ object containing a key-value pair for each predicted property.\subsection{AFLOW: Database Organization}
\label{database_organization}

The \AFLOW\ data repository \cite{curtarolo:art75} is organized into project, set and calculation
layers as illustrated in Figure \ref{fig:aflow_webpage}(c). 
At the project layer, the calculations are divided into different catalogs 
based on the origin and composition of the entries \cite{curtarolo:art92, curtarolo:art128}.
Within each catalog, entries are grouped into sets based on shared lattice type or alloy system.
The entries within each set contain the results of DFT calculated properties for particular structures.

The \AFLOW-ICSD catalog contains the DFT calculated properties for over 57,000 experimentally observed materials listed in the
\underline{I}norganic \underline{C}rystal \underline{S}tructure \underline{D}atabase (ICSD) \cite{ICSD, ICSD1}. 
Internally, this catalog is organized by lattice type, and then by individual materials entry. 
Since the materials in this catalog are already known to exist, the primary interest is in accurately calculating
electronic structure and thermo-mechanical properties. 
Therefore, calculations for this catalog are generally performed using the Hubbard $U$ correction to the DFT 
exchange-correlation functional \cite{Liechtenstein1995, Dudarev1998} where appropriate, using a set of standardized $U$ values \cite{curtarolo:art104}. 
Within this catalog, entries are grouped by Bravais lattice type 
into 14 sets: 
``{\small BCC}'',
 ``{\small BCT}'', 
``{\small CUB}'', 
``{\small FCC}'', 
``{\small HEX'}'', 
``{\small MCL}''', 
``{\small MCLC}''',
 ``{\small ORC}''', 
``{\small ORCC}''', 
``{\small ORCF}''', 
``{\small ORCI}''', 
``{\small RHL}''', 
``{\small TET}''' and
``{\small TRI}'''. 
The name of each materials entry is generated using the format \verb|<composition>_ICSD_<ICSD number>|.

The entries in the other catalogs, such as ``LIB1'', ``LIB2'', and ``LIB3'', 
are generated by decorating crystal structure prototypes to predict new hypothetical compounds, and contain unary, binary, and ternary materials, respectively. 
Additional catalogs, ``LIB4'', ``LIB5'', and ``LIB6'', are currently being generated for quaternary, quinary, and hexenary materials.
Within each catalog, the entries are grouped by element and exchange-correlation functional in the case of ``LIB1'', 
and by alloy system in the cases of ``LIB2'' and ``LIB3''. 
``LIB1'' contains a total of 2784 entries, while ``LIB2'' currently has 329,181 entries and ``LIB3'' has over 1.3 million.
Within each alloy system, the individual materials entries are named according the relevant crystal prototype. 
For these catalogs, the emphasis is on the discovery of new thermodynamically stable or metastable materials, 
and on their use to generate the thermodynamic density of states for the prediction of the formation of disordered 
materials such as metallic glasses \cite{curtarolo:art112} or high entropy alloys \cite{Lederer_HEA_2017}. 
Therefore, calculations in these catalogs are performed using the GGA-PBE exchange-correlation functional \cite{PBE} without Hubbard $U$ corrections \cite{curtarolo:art104} 
so as to produce consistent energy differences, enabling the calculation of accurate  formation enthalpies.

\subsection{AFLOW: Database Properties}

\begin{figure*}
  \includegraphics[width=0.98\textwidth]{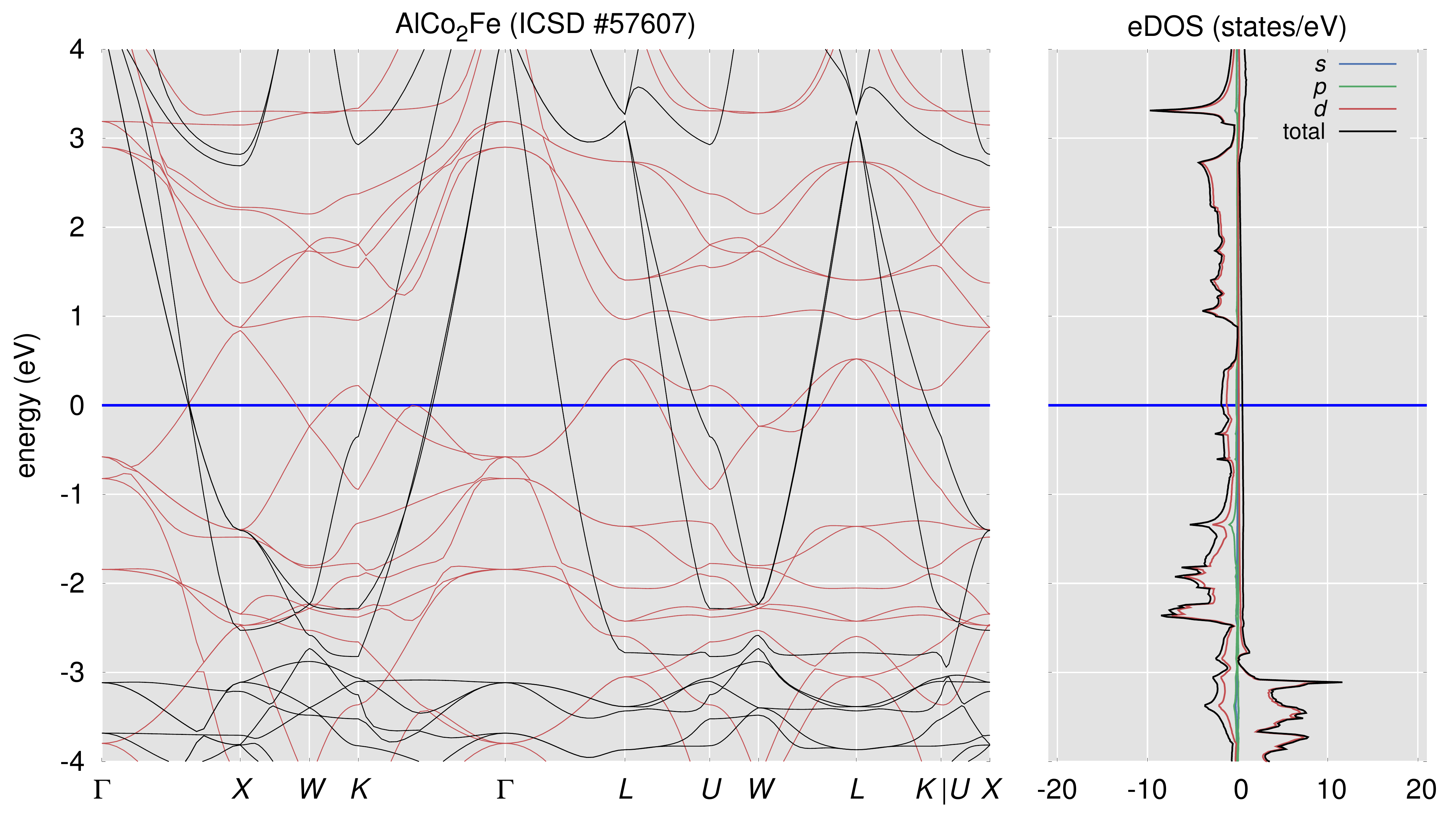}
  \vspace{-4mm}
  \caption{\small \textbf{Example band structure and density of states images automatically generated and 
    served through the \AFLOW.org data repository.}
    The structure highlighted is AlCo$_{2}$Fe (ICSD \#57607): \url{http://aflow.org/material.php?id=Al1Co2Fe1_ICSD_57607}.
    The results of the spin-polarized calculation are differentiated by: color on the band structure plot
    (black/red for majority/minority spin), and sign on the density of states plot (positive/negative for majority/minority spin).
    The band structure is calculated following the \AFLOW\ Standard path of high-symmetry \textbf{k}-points~\cite{curtarolo:art58}.
  }
  \label{fig:band_structure}
\end{figure*}

Materials properties within the \AFLOW\ repository \cite{curtarolo:art75} are indexed as keyword-value pairs which are 
programmatically accessible via the \AFLOW\ Data \RESTAPI\ \cite{curtarolo:art92} and programmatically searchable via
the \AFLUX\ Search-{\small API} \cite{curtarolo:art128}. 
Search filters for these properties are also available in the advanced search application of the \verb|aflow.org| web portal as highlighted by 
the dashed red rectangle in Figure \ref{fig:aflow_webpage}(b),  where they are grouped into chemistry ({\it e.g.}, chemical species, stoichiometry), 
crystal ({\it e.g.}, space group, Bravais lattice type), electronic ({\it e.g.} band gaps), 
thermodynamic (energetic and thermal properties, {\it e.g.} formation enthalpies and Debye temperatures), magnetic, scintillation, 
mechanical (elastic moduli and pressure-related properties), and calculation ({\it e.g.}, {\bf k}-point mesh, \AFLOW\ version) parameters. 
In total there are in excess of 170 million individual materials properties indexed in the \AFLOW\ database ($\sim 100$ per materials entry). 
Lists of the keywords corresponding to the materials properties are provided in 
Refs. \onlinecite{curtarolo:art92,curtarolo:art128, curtarolo:art115}.

Systems for which the ``STATIC'' and ``BANDS'' calculations have been performed are supplemented 
with automatically generated images of the density of states, projected density of states, and band structure.
Both low (PNG) and high (EPS) quality variants of the images are available for download.
An example band structure and density of states image is displayed in Figure~\ref{fig:band_structure}.

\subsection{AFLOW: Data {\small REST-API}}

The full data set generated by the high-throughput \AFLOW\ process \cite{curtarolo:art75} is backed
by a disk store of (at this time) over 12 TB of input criteria, calculated results, and derivative output.  
The backing store is exposed via the \AFLOW\ Data \RESTAPI\ \cite{curtarolo:art75} in a hierarchical organization. 
This direct exposure of our results not only grants the end user a high degree of utility via direct access,
but more importantly, guarantees data provenance that promotes reproducibility. 
The hierarchy of the \AFLOW\ Data \RESTAPI\ categorizes this abundance of information into meaningful high-level 
classifications allowing for exploration of self-similar materials that are related by stoichiometric and/or
crystallographic properties. 
Once a selection of materials has been determined, the full range of available properties and procedural data are retrievable. 

The organizational hierarchy of both the underlying data store and the \RESTAPI\ is project dependent, as described 
in Section \ref{database_organization}. 
Each project is equivalent to one of the catalogs listed in Section \ref{database_organization}, 
and in the \RESTAPI\ are denoted by the project layers ``ICSD\_WEB'', ``LIB1\_RAW'', ``LIB2\_RAW'', and ``LIB3\_RAW''. 
Each project layer contains multiple set layers, which correspond to specific alloy systems in the case of ``LIB1\_RAW'', ``LIB2\_RAW'', and ``LIB3\_RAW''. 
For instance: \verb|http://aflowlib.duke.edu/AFLOWDATA/LIB2_RAW/| exposes the set layer for
binary entries, where each set corresponds to different binary alloy system, allowing for pairwise atomic species examination. 
Within each set is the entry layer, consisting of decorated structural prototypes which provide stoichiometric and structural variation for each alloy system. 
Each entry contains the calculated results for a particular structure and composition, organized as keyword-value pairs.
The calculated values of thermodynamical, mechanical, electronic, magnetic, chemical and crystallographic properties can
be directly accessed by querying a \underline{U}niform \underline{R}esource \underline{I}dentifier (\URI) of the form \verb|<server>/<project>/<set>/<entry>/?<keyword>|, 
where \verb|<server>| is \verb|http://aflowlib.duke.edu/AFLOWDATA|, \verb|<project>| is the appropriate
project layer, \verb|<set>| is the alloy system, \verb|<entry>| is the structural prototype, and \verb|<keyword>| 
corresponds to the materials property of interest. 
A full description of the \RESTAPI\ keywords is provided in Ref. \onlinecite{curtarolo:art92},
along with additions in the appendices of Refs. \onlinecite{curtarolo:art128, curtarolo:art115}.

The ability to explore related entries predicated on a multitude of properties leads directly to novel materials discovery and use. 
The \AFLOW\ Data \RESTAPI\ disseminates our methods and results, without restriction, to a
global research audience in order to promote scientific and engineering advancement.

\subsection{AFLUX: Search-{\small API}}

The \underline{A}utomatic \underline{F}low of \LUX\ or \AFLUX\ Search-{\small API} \cite{curtarolo:art128} is a human usable remote data
search {\small API}. \LUX\ is designed to be a domain agnostic solution to the
outstanding problem of programmatically searching remote data that typically is
either exposed via a capriciously limited utility or requires {\it a-priori}
knowledge of the internal organization of the remote repository.  The \LUX\ query
concept flattens the exposed data, while simultaneously providing arbitrarily
complex query capability, allowing an end user full freedom in constraining the
requested data. \LUX\ is designed to operate in the nearly ubiquitous
web \URI\ context while minimizing any potentially conflicting interactions with existing
\URI\ functionality. 

\AFLUX\ is the domain specific implementation of \LUX\ and is available at the
\verb|<AFLUX-URI>|: \verb|http://aflowlib.duke.edu/search/API/?|.
At this time, the \AFLUX\ {\small API} freely exposes over 170 million keyword-value
properties without any requirements or restrictions on the end user. 
Specific properties and compositions can be searched for by appending
the appropriate keywords to the \verb|<AFLUX-URI>|. 
Search results can be restricted by including specific values or value ranges in parentheses
after the appropriate keyword. 
For example, a search can be restricted to entries that contain both of the elements Na and Cl, 
and have a calculated electronic band gap in excess of 1.0eV, by including the
search parameters \verb|species(Na,Cl),Egap(1.0*)| in the query part of the \URI. 
In \LUX, ``,'' corresponds to the logical \verb|AND| operator, and ``*'' is the \verb|loose| operator 
which extends the search to entries in a specified value range. 
If no parameters are provided for a particular keyword, then the values of that property are 
returned for all entries which satisfy the remaining search criteria.
A full list of all \LUX\ logical operators can be found in Ref. \onlinecite{curtarolo:art128},
along with descriptions of their functionality and appropriate usage.

In addition to materials properties keywords, \LUX\ also accepts
directives, which behave as pseudo property keywords. They are used to provide additional information on \LUX\ 
usage, and control the format and quantity of the returned data. Note that any directives
included in a search query must come after all of the materials properties keywords.
In particular, the \verb|schema| directive can be used to retrieve the 
most current and canonical list of keywords using the \AFLUX\ summons:
\verb|<AFLUX-URI>schema,format(json)|.

\subsection{AFLOW: Data Quality Control}
\label{sec:valver}

Data quality control, including validation of methodologies and verification of calculated data, is vital when constructing large databases such as the \AFLOW\ repository \cite{curtarolo:art75}
in order to guarantee the reliability of the results. 
Methodological validation involves quantifying the accuracy of calculation models with respect to experiment, while 
data verification includes checking the robustness of calculation parameters and the satisfaction of convergence criteria.

Physical models incorporated into the \AFLOW\ framework are validated by comparison to benchmark sets of experimental data. This helps determine the predictive accuracy of the methods for real materials, as well as the regimes in which they are reliable. For example, the \AEL\
and \AGL\ modules were validated by comparison to a benchmark set of $\sim75$ experimentally well characterized compounds of various structural types \cite{curtarolo:art115, curtarolo:art96}, and the accuracy was quantified by the Pearson and Spearman correlations, and the root-mean-square deviations. Similar validation analyses were performed for the \QHAAPL\ \cite{curtarolo:art114, curtarolo:art119} and \AAPL\ methods \cite{curtarolo:art125}, as well as 
the property labelled materials fragments machine learning model \cite{curtarolo:art124}.

The \AFLOWPOCC\ methodology has been validated by comparing the band gap as a function of composition for ZnS$_{1-x}$Se$_x$ and Mg$_x$Zn$_{1-x}$O, and the magnetic moment per atom as a function of composition for Fe$_{1-x}$Cu$_x$, to experimental values \cite{curtarolo:art110}.  

The ACBN0 functional \cite{curtarolo:art93}, implemented in \AFLOW$\pi$ (see Section \ref{sec:aflowpi}) and PAOFLOW (see Section \ref{sec:paoflow})
 has been validated by comparing the lattice parameters, bulk moduli, electronic band gaps, phonon modes, high frequency dielectric
constants and Born effective changes it produces to the experimentally measured values for the Zn and Cd chalcogenides \cite{curtarolo:art103}.

The convergence of both the charge density optimization and the ionic structural relaxation are automatically verified for all \AFLOW\ calculations prior to incorporation into the data repository. This includes, for example, checking that the charge density has converged in accordance with the \AFLOW\ standard settings \cite{curtarolo:art104}, and verifying the relaxation of the cell size and shape by ensuring that all elements of the stress tensor are less than 10kB. The convergence level for any individual calculation can be verified by querying appropriate keywords for the stress tensor: \verb|stress_tensor|, Pulay stress: \verb|Pulay_stress|, residual external pressure on the relaxed cell: \verb|pressure_residual|,  and the $\delta E$ value for the final electronic convergence step: \verb|delta_electronic_energy_convergence|, using the \AFLOW\ Data {\small REST-API} \cite{curtarolo:art92} or the \AFLUX\ Search-{\small API} \cite{curtarolo:art128}. Initial calculation parameters can similarly be obtained using the appropriate keywords, such as the k-point grid: \verb|kpoints|, or the electronic energy convergence threshold: \verb|delta_electronic_energy_threshold|.

\section*{Conclusion}
The \AFLOW\ Fleet for computational materials design automates first principles calculations of materials properties. 
\AFLOW\ incorporates a wide range of different modules, including applications for symmetry and thermodynamic stability analysis, 
generation of ordered and disordered materials structures, and calculation of thermo-mechanical properties, in a single
integrated framework.
\AFLOWpi\ is a versatile minimalist framework that includes tools for projection onto pseudo-atomic orbitals (PAO) and
the self-consistent calculation of Hubbard $U$ corrections using ACBN0. 
\PAOFLOW\ generates tight-binding hamiltonians which reproduce the electronic structure calculated using first principles methods,
facilitating the rapid calculation of electronic and magnetic properties such as transport coefficients and the Berry curvature.
All results are stored in, and disseminated through, the \AFLOW\ data repository, which is available online at \verb|aflow.org|, and is programmatically
accessible via the \AFLOW\ Data \RESTAPI\ and the \AFLUX\ Search-{\small API}.

\section*{Acknowledgements}
The authors acknowledge support from 
{\small DOD-ONR} (N00014-13-1-0030, N00014-13-1-0635, N00014-17-1-2090, N00014-16-1-2781, N00014-15-1-2583, N00014-15-1-2266), 
{\small DOE} (DE-AC02-05CH11231, specifically {\small BES} Grant \# EDCBEE), and the Duke University Center for Materials Genomics.  
SC acknowledges support by the Alexander von Humboldt-Foundation - Max Plank Society (Fritz-Haber-Institut der Max-Planck-Gesellschaft, Berlin-Dahlem, Germany).
CO acknowledges support from the National Science Foundation Graduate Research Fellowship under Grant No. DGF-1106401.  
\AFLOW\ calculations were performed at the Duke University Center for Materials Genomics and at the Fulton Supercomputer Lab - Brigham Young University.
The authors thank Amir Natan, Matthias Scheffler, Luca Ghiringhelli, Kenneth Vecchio, Don Brenner and Jon-Paul Maria for helpful discussions.


\section*{Corresponding authors}
Correspondence to Cormac Toher (\verb|toherc@gmail.com|) and/or Stefano Curtarolo (\verb|stefano@duke.edu|).

\newcommand{\Ozolins}{Ozoli\c{n}\v{s}}


\begin{thebibliography}{10}
\expandafter\ifx\csname urlstyle\endcsname\relax
  \providecommand{\doi}[1]{doi:\discretionary{}{}{}#1}\else
  \providecommand{\doi}{doi:\discretionary{}{}{}\begingroup
  \urlstyle{rm}\Url}\fi
\providecommand{\selectlanguage}[1]{\relax}
\providecommand{\bibAnnoteFile}[1]{%
  \IfFileExists{#1}{\begin{quotation}\noindent\textsc{Key:} #1\\
  \textsc{Annotation:}\ \input{#1}\end{quotation}}{}}
\providecommand{\bibAnnote}[2]{%
  \begin{quotation}\noindent\textsc{Key:} #1\\
  \textsc{Annotation:}\ #2\end{quotation}}

\bibitem{nmatHT}
S.~Curtarolo, G.~L.~W. Hart, M.~{Buongiorno~Nardelli}, N.~Mingo, S.~Sanvito,
  and O.~Levy, \emph{The high-throughput highway to computational materials
  design}, Nat.\ Mater. \textbf{12}, 191--201 (2013).
\input{nmatHT}

\bibitem{curtarolo:art65}
S.~Curtarolo, W.~Setyawan, G.~L.~W. Hart, M.~Jahn\'{a}tek, R.~V. Chepulskii,
  R.~H. Taylor, S.~Wang, J.~Xue, K.~Yang, O.~Levy, M.~J. Mehl, H.~T. Stokes,
  D.~O. Demchenko, and D.~Morgan, \emph{{AFLOW}: An automatic framework for
  high-throughput materials discovery}, Comput.\ Mater.\ Sci. \textbf{58},
  218--226 (2012).
\input{curtarolo:art65}

\bibitem{curtarolo:art127}
A.~R. Supka, T.~E. Lyons, L.~S.~I. Liyanage, P.~{D'{A}mico},
  R.~{Al~Rahal~Al~Orabi}, S.~Mahatara, P.~Gopal, C.~Toher, D.~Ceresoli,
  A.~Calzolari, S.~Curtarolo, M.~{Buongiorno~Nardelli}, and M.~Fornari,
  \emph{{\small AFLOW}$\pi$: A minimalist approach to high-throughput {\it ab
  initio} calculations including the generation of tight-binding hamiltonians},
  Comput.\ Mater.\ Sci. \textbf{136}, 76--84 (2017).
\input{curtarolo:art127}

\bibitem{paoflow}
M.~{Buongiorno Nardelli}, F.~T. Cerasoli, M.~Costa, S.~Curtarolo, R.~D.
  Gennaro, M.~Fornari, L.~Liyanage, A.~R. Supka, and H.~Wang, \emph{{PAOFLOW}:
  A utility to construct and operate on ab initio Hamiltonians from the
  {P}rojections of electronic wavefunctions on {A}tomic {O}rbital bases,
  including characterization of topological materials}, Comput.\ Mater.\ Sci.
  (2017).
\input{paoflow}

\bibitem{curtarolo:art75}
S.~Curtarolo, W.~Setyawan, S.~Wang, J.~Xue, K.~Yang, R.~H. Taylor, L.~J.
  Nelson, G.~L.~W. Hart, S.~Sanvito, M.~{Buongiorno~Nardelli}, N.~Mingo, and
  O.~Levy, \emph{{AFLOWLIB.ORG}: A distributed materials properties repository
  from high-throughput {\it ab initio} calculations}, Comput.\ Mater.\ Sci.
  \textbf{58}, 227--235 (2012).
\input{curtarolo:art75}

\bibitem{curtarolo:art92}
R.~H. Taylor, F.~Rose, C.~Toher, O.~Levy, K.~Yang, M.~{Buongiorno~Nardelli},
  and S.~Curtarolo, \emph{A {REST}ful {API} for exchanging materials data in
  the {AFLOWLIB}.org consortium}, Comput.\ Mater.\ Sci. \textbf{93}, 178--192
  (2014).
\input{curtarolo:art92}

\bibitem{curtarolo:art128}
F.~Rose, C.~Toher, E.~Gossett, C.~Oses, M.~{Buongiorno~Nardelli}, M.~Fornari,
  and S.~Curtarolo, \emph{{AFLUX}: The {LUX} materials search {API} for the
  {AFLOW} data repositories}, Comput.\ Mater.\ Sci. \textbf{137}, 362--370
  (2017).
\input{curtarolo:art128}

\bibitem{APL_Mater_Jain2013}
A.~Jain, S.~P. Ong, G.~Hautier, W.~Chen, W.~D. Richards, S.~Dacek, S.~Cholia,
  D.~Gunter, D.~Skinner, G.~Ceder, and K.~A. Persson, \emph{{Commentary: The
  Materials Project: A materials genome approach to accelerating materials
  innovation}}, APL\ Mater. \textbf{1}, 011002 (2013).
\input{APL_Mater_Jain2013}

\bibitem{nomad}
M.~Scheffler, C.~Draxl, and {Computer Center of the Max-Planck Society,
  Garching}, \emph{The {NoMaD} Repository}, http://nomad-repository.eu (2014).
\input{nomad}

\bibitem{oqmd.org}
J.~E. Saal, S.~Kirklin, M.~Aykol, B.~Meredig, and C.~Wolverton, \emph{Materials
  Design and Discovery with High-Throughput Density Functional Theory: The
  {O}pen {Q}uantum {M}aterials {D}atabase ({OQMD})}, JOM \textbf{65},
  1501--1509 (2013).
\input{oqmd.org}

\bibitem{cmr_repository}
D.~D. Landis, J.~Hummelsh{\o}j, S.~Nestorov, J.~Greeley, M.~Du{\l}ak,
  T.~Bligaard, J.~K. N{\o}rskov, and K.~W. Jacobsen, \emph{The Computational
  Materials Repository}, Comput.\ Sci.\ Eng. \textbf{14}, 51--57 (2012).
\input{cmr_repository}

\bibitem{Pizzi_AiiDA_2016}
G.~Pizzi, A.~Cepellotti, R.~Sabatini, N.~Marzari, and B.~Kozinsky,
  \emph{{AiiDA}: automated interactive infrastructure and database for
  computational science}, Comput.\ Mater.\ Sci. \textbf{111}, 218--230 (2016).
\input{Pizzi_AiiDA_2016}

\bibitem{CMS_Ong2012b}
S.~P. Ong, W.~D. Richards, A.~Jain, G.~Hautier, M.~Kocher, S.~Cholia,
  D.~Gunter, V.~L. Chevrier, K.~A. Persson, and G.~Ceder, \emph{{Python
  Materials Genomics (pymatgen): A robust, open-source python library for
  materials analysis}}, Comput.\ Mater.\ Sci. \textbf{68}, 314--319 (2013).
\input{CMS_Ong2012b}

\bibitem{ase}
S.~R. Bahn and K.~W. Jacobsen, \emph{An object-oriented scripting interface to
  a legacy electronic structure code}, Comput.\ Sci.\ Eng. \textbf{4}, 56--66
  (2002).
\input{ase}

\bibitem{kresse_vasp}
G.~Kresse and J.~Hafner, \emph{{\it Ab initio} molecular dynamics for liquid
  metals}, Phys.\ Rev.\ B \textbf{47}, 558--561 (1993).
\input{kresse_vasp}

\bibitem{vasp}
G.~Kresse and J.~Furthm\"{u}ller, \emph{Efficient iterative schemes for {\it ab
  initio} total-energy calculations using a plane-wave basis set}, Phys.\ Rev.\
  B \textbf{54}, 11169--11186 (1996).
\input{vasp}

\bibitem{qe}
P.~Giannozzi, S.~Baroni, N.~Bonini, M.~Calandra, R.~Car, C.~Cavazzoni,
  D.~Ceresoli, G.~L. Chiarotti, M.~Cococcioni, I.~Dabo, A.~{Dal Corso}, S.~{de
  Gironcoli}, S.~Fabris, G.~Fratesi, R.~Gebauer, U.~Gerstmann, C.~Gougoussis,
  A.~Kokalj, M.~Lazzeri, L.~Martin-Samos, N.~Marzari, F.~Mauri, R.~Mazzarello,
  S.~Paolini, A.~Pasquarello, L.~Paulatto, C.~Sbraccia, S.~Scandolo,
  G.~Sclauzero, A.~P. Seitsonen, A.~Smogunov, P.~Umari, and R.~M. Wentzcovitch,
  \emph{QUANTUM ESPRESSO: a modular and open-source software project for
  quantum simulations of materials}, J.\ Phys.:\ Condens.\ Matter \textbf{21},
  395502 (2009).
\input{qe}

\bibitem{Giannozzi:2017io}
P.~Giannozzi, O.~Andreussi, T.~Brumme, O.~Bunau, M.~{Buongiorno~Nardelli},
  M.~Calandra, R.~Car, C.~Cavazzoni, D.~Ceresoli, M.~Cococcioni, N.~Colonna,
  I.~Carnimeo, A.~{Dal~Corso}, S.~{de~Gironcoli}, P.~Delugas, R.~A.
  {DiStasio~Jr.}, A.~Ferretti, A.~Floris, G.~Fratesi, G.~Fugallo, R.~Gebauer,
  U.~Gerstmann, F.~Giustino, T.~Gorni, J.~Jia, M.~Kawamura, H.-Y. Ko,
  A.~Kokalj, E.~K\"{u}\c{c}\"{u}kbenli, M.~Lazzeri, M.~Marsili, N.~Marzari,
  F.~Mauri, N.~L. Nguyen, H.-V. Nguyen, A.~{Otero-de-la-Roza}, L.~Paulatto,
  S.~Ponc{\'e}, D.~Rocca, R.~Sabatini, B.~Santra, M.~Schlipf, A.~P. Seitsonen,
  A.~Smogunov, I.~Timrov, T.~Thonhauser, P.~Umari, N.~Vast, X.~Wu, and
  S.~Baroni, \emph{Advanced capabilities for materials modelling with Quantum
  {ESPRESSO}}, J.\ Phys.:\ Condens.\ Matter \textbf{29}, 465901 (2017).
\input{Giannozzi:2017io}

\bibitem{curtarolo:art112}
E.~Perim, D.~Lee, Y.~Liu, C.~Toher, P.~Gong, Y.~{Li}, W.~N. Simmons, O.~Levy,
  J.~J. Vlassak, J.~Schroers, and S.~Curtarolo, \emph{Spectral descriptors for
  bulk metallic glasses based on the thermodynamics of competing crystalline
  phases}, Nat.\ Commun. \textbf{7}, 12315 (2016).
\input{curtarolo:art112}

\bibitem{curtarolo:art109}
S.~Sanvito, C.~Oses, J.~Xue, A.~Tiwari, M.~Zic, T.~Archer, P.~Tozman,
  M.~Venkatesan, J.~M.~D. Coey, and S.~Curtarolo, \emph{Accelerated discovery
  of new magnets in the {H}eusler alloy family}, Sci.\ Adv. \textbf{3},
  e1602241 (2017).
\input{curtarolo:art109}

\bibitem{curtarolo:art113}
C.~Nyshadham, C.~Oses, J.~E. Hansen, I.~Takeuchi, S.~Curtarolo, and G.~L.~W.
  Hart, \emph{A computational high-throughput search for new ternary
  superalloys}, Acta\ Mater. \textbf{122}, 438--447 (2017).
\input{curtarolo:art113}

\bibitem{curtarolo:art106}
S.~Barzilai, C.~Toher, S.~Curtarolo, and O.~Levy, \emph{Evaluation of the
  tantalum-titanium phase diagram from {\it ab-initio} calculations}, Acta\
  Mater. \textbf{120}, 255--263 (2016).
\input{curtarolo:art106}

\bibitem{curtarolo:art117}
S.~Barzilai, C.~Toher, S.~Curtarolo, and O.~Levy, \emph{The effect of lattice
  stability determination on the computational phase diagrams of intermetallic
  alloys}, J.\ Alloys\ Compd. \textbf{728}, 314--321 (2017).
\input{curtarolo:art117}

\bibitem{Lederer_HEA_2017}
Y.~Lederer, C.~Toher, K.~S. Vecchio, and S.~Curtarolo, \emph{The search for
  high entropy alloys: a high-throughput {\it ab-initio} approach}, submitted
  arxiv.org/1711.03426  (2017).
\input{Lederer_HEA_2017}

\bibitem{curtarolo:art49}
O.~Levy, G.~L.~W. Hart, and S.~Curtarolo, \emph{Uncovering Compounds by Synergy
  of Cluster Expansion and High-Throughput Methods}, J.\ Am.\ Chem.\ Soc.
  \textbf{132}, 4830--4833 (2010).
\input{curtarolo:art49}

\bibitem{curtarolo:art51}
O.~Levy, G.~L.~W. Hart, and S.~Curtarolo, \emph{Hafnium Binary Alloys from
  Experiments and First Principles}, Acta\ Mater. \textbf{58}, 2887--2897
  (2010).
\input{curtarolo:art51}

\bibitem{curtarolo:art53}
O.~Levy, R.~V. Chepulskii, G.~L.~W. Hart, and S.~Curtarolo, \emph{The New face
  of {Rh}odium Alloys: Revealing Ordered Structures from First Principles}, J.\
  Am.\ Chem.\ Soc. \textbf{132}, 833--837 (2010).
\input{curtarolo:art53}

\bibitem{curtarolo:art126}
S.~Barzilai, C.~Toher, S.~Curtarolo, and O.~Levy, \emph{Molybdenum-titanium
  phase diagram evaluated from \textit{ab initio} calculations}, Phys.\ Rev.\
  Mater. \textbf{1}, 023604 (2017).
\input{curtarolo:art126}

\bibitem{curtarolo:art94}
O.~Isayev, D.~Fourches, E.~N. Muratov, C.~Oses, K.~Rasch, A.~Tropsha, and
  S.~Curtarolo, \emph{Materials Cartography: Representing and Mining Materials
  Space Using Structural and Electronic Fingerprints}, Chem.\ Mater.
  \textbf{27}, 735--743 (2015).
\input{curtarolo:art94}

\bibitem{curtarolo:art124}
O.~Isayev, C.~Oses, C.~Toher, E.~Gossett, S.~Curtarolo, and A.~Tropsha,
  \emph{Universal fragment descriptors for predicting electronic properties of
  inorganic crystals}, Nat.\ Commun. \textbf{8}, 15679 (2017).
\input{curtarolo:art124}

\bibitem{curtarolo:art104}
C.~E. Calderon, J.~J. Plata, C.~Toher, C.~Oses, O.~Levy, M.~Fornari, A.~Natan,
  M.~J. Mehl, G.~L.~W. Hart, M.~{Buongiorno~Nardelli}, and S.~Curtarolo,
  \emph{The {AFLOW} standard for high-throughput materials science
  calculations}, Comput.\ Mater.\ Sci. \textbf{108 Part A}, 233--238 (2015).
\input{curtarolo:art104}

\bibitem{PAW}
P.~E. Bl\"{o}chl, \emph{Projector augmented-wave method}, Phys.\ Rev.\ B
  \textbf{50}, 17953--17979 (1994).
\input{PAW}

\bibitem{PBE}
J.~P. Perdew, K.~Burke, and M.~Ernzerhof, \emph{Generalized Gradient
  Approximation Made Simple}, Phys.\ Rev.\ Lett. \textbf{77}, 3865--3868
  (1996).
\input{PBE}

\bibitem{curtarolo:art58}
W.~Setyawan and S.~Curtarolo, \emph{High-throughput electronic band structure
  calculations: Challenges and tools}, Comput.\ Mater.\ Sci. \textbf{49},
  299--312 (2010).
\input{curtarolo:art58}

\bibitem{curtarolo:art96}
C.~Toher, J.~J. Plata, O.~Levy, M.~{de~Jong}, M.~D. Asta,
  M.~{Buongiorno~Nardelli}, and S.~Curtarolo, \emph{High-throughput
  computational screening of thermal conductivity, {D}ebye temperature, and
  {G}r\"{u}neisen parameter using a quasiharmonic {D}ebye model}, Phys.\ Rev.\
  B \textbf{90}, 174107 (2014).
\input{curtarolo:art96}

\bibitem{curtarolo:art115}
C.~Toher, C.~Oses, J.~J. Plata, D.~Hicks, F.~Rose, O.~Levy, M.~{de Jong}, M.~D.
  Asta, M.~Fornari, M.~{Buongiorno~Nardelli}, and S.~Curtarolo, \emph{Combining
  the {AFLOW} {GIBBS} and Elastic Libraries to efficiently and robustly screen
  thermomechanical properties of solids}, Phys.\ Rev.\ Mater. \textbf{1},
  015401 (2017).
\input{curtarolo:art115}

\bibitem{curtarolo:art114}
P.~Nath, J.~J. Plata, D.~Usanmaz, R.~{Al~Rahal~Al~Orabi}, M.~Fornari,
  M.~{Buongiorno~Nardelli}, C.~Toher, and S.~Curtarolo, \emph{High-throughput
  prediction of finite-temperature properties using the quasi-harmonic
  approximation}, Comput.\ Mater.\ Sci. \textbf{125}, 82--91 (2016).
\input{curtarolo:art114}

\bibitem{curtarolo:art125}
J.~J. Plata, P.~Nath, D.~Usanmaz, J.~Carrete, C.~Toher, M.~Fornari,
  M.~{Buongiorno~Nardelli}, and S.~Curtarolo, \emph{An efficient and accurate
  framework for calculating lattice thermal conductivity of solids:
  {AFLOW}-{AAPL} {Au}tomatic {A}nharmonic {P}honon {Li}brary}, NPJ\ Comput.\
  Mater. \textbf{3}, 45 (2017).
\input{curtarolo:art125}

\bibitem{curtarolo:art110}
K.~Yang, C.~Oses, and S.~Curtarolo, \emph{Modeling Off-Stoichiometry Materials
  with a High-Throughput {\it Ab-Initio} Approach}, Chem.\ Mater. \textbf{28},
  6484--6492 (2016).
\input{curtarolo:art110}

\bibitem{curtarolo:art121}
M.~J. Mehl, D.~Hicks, C.~Toher, O.~Levy, R.~M. Hanson, G.~L.~W. Hart, and
  S.~Curtarolo, \emph{The {AFLOW} Library of Crystallographic Prototypes: Part
  1}, Comput.\ Mater.\ Sci. \textbf{136}, S1--S828 (2017).
\input{curtarolo:art121}

\bibitem{gonze:abinit}
X.~Gonze, J.~M. Beuken, R.~Caracas, F.~Detraux, M.~Fuchs, G.~M. Rignanese,
  L.~Sindic, M.~Verstraete, G.~Zerah, F.~Jollet, M.~Torrent, A.~Roy, M.~Mikami,
  P.~Ghosez, J.~Y. Raty, and D.~Allan, \emph{First-principles computation of
  material properties: the {ABINIT} software project}, Comput.\ Mater.\ Sci.
  \textbf{25}, 478--492 (2002).
\input{gonze:abinit}

\bibitem{blum:fhi-aims}
V.~Blum, R.~Gehrke, F.~Hanke, P.~Havu, V.~Havu, X.~Ren, K.~Reuter, and
  M.~Scheffler, \emph{Ab initio molecular simulations with numeric
  atom-centered orbitals}, Comput.\ Phys.\ Commun. \textbf{180}, 2175--2196
  (2009).
\input{blum:fhi-aims}

\bibitem{aflowsym_2017}
D.~Hicks, C.~Oses, R.~H. Taylor, E.~Gossett, G.~Gomez, C.~Toher, and
  S.~Curtarolo, \emph{{AFLOW-SYM}: Platform for the complete, automatic and
  self-consistent symmetry analysis of crystals}, submitted  (2017).
\input{aflowsym_2017}

\bibitem{qhull}
C.~B. Barber, D.~P. Dobkin, and H.~Huhdanpaa, \emph{The quickhull algorithm for
  convex hulls}, ACM Trans. Math. Soft. \textbf{22}, 469--483 (1996).
\input{qhull}

\bibitem{gus_enum1}
G.~L.~W. Hart and R.~W. Forcade, \emph{Algorithm for generating derivative
  structures}, Phys.\ Rev.\ B \textbf{77}, 224115 (2008).
\input{gus_enum1}

\bibitem{Rappe_1992_JCAS_UFF}
A.~K. Rappe, C.~J. Casewit, K.~S. Colwell, W.~A. Goddard, and W.~M. Skiff,
  \emph{{UFF}, a full periodic table force field for molecular mechanics and
  molecular dynamics simulations}, J.\ Am.\ Chem.\ Soc. \textbf{114},
  10024--10035 (1992).
\input{Rappe_1992_JCAS_UFF}

\bibitem{curtarolo:art100}
M.~{de~Jong}, W.~Chen, T.~Angsten, A.~Jain, R.~Notestine, A.~Gamst, M.~Sluiter,
  C.~K. Ande, S.~{van~der~Zwaag}, J.~J. Plata, C.~Toher, S.~Curtarolo,
  G.~Ceder, K.~A. Persson, and M.~D. Asta, \emph{Charting the Complete Elastic
  properties of Inorganic Crystalline Compounds}, Sci.\ Data \textbf{2}, 150009
  (2015).
\input{curtarolo:art100}

\bibitem{Poirier_Earth_Interior_2000}
J.-P. Poirier, \emph{Introduction to the Physics of the Earth’s Interior}
  (Cambridge University Press, 2000), 2nd edn.
\input{Poirier_Earth_Interior_2000}

\bibitem{Blanco_CPC_GIBBS_2004}
M.~A. Blanco, E.~Francisco, and V.~Lua{\~n}a, \emph{GIBBS: isothermal-isobaric
  thermodynamics of solids from energy curves using a quasi-harmonic Debye
  model}, Comput.\ Phys.\ Commun. \textbf{158}, 57--72 (2004).
\input{Blanco_CPC_GIBBS_2004}

\bibitem{Blanco_jmolstrthch_1996}
M.~A. Blanco, A.~M. Pend\'{a}s, E.~Francisco, J.~M. Recio, and R.~Franco,
  \emph{Thermodynamical properties of solids from microscopic theory:
  Applications to MgF$_2$ and Al$_2$O$_3$}, J.\ Mol.\ Struct.,\ Theochem
  \textbf{368}, 245--255 (1996).
\input{Blanco_jmolstrthch_1996}

\bibitem{Maradudin1971}
A.~A. Maradudin, E.~W. Montroll, G.~H. Weiss, and I.~P. Ipatova, \emph{Theory
  of Lattice Dynamics in the Harmonic Approximation} (Academic Press, New York,
  1971).
\input{Maradudin1971}

\bibitem{curtarolo:art119}
P.~Nath, J.~J. Plata, D.~Usanmaz, C.~Toher, M.~Fornari,
  M.~{Buongiorno~Nardelli}, and S.~Curtarolo, \emph{High throughput
  combinatorial method for fast and robust prediction of lattice thermal
  conductivity}, Scr.\ Mater. \textbf{129}, 88--93 (2017).
\input{curtarolo:art119}

\bibitem{Schmalzl_prb_2003}
K.~Schmalzl, D.~Strauch, and H.~Schober, \emph{Lattice-dynamical and
  ground-state properties of CaF$_2$ studied by inelastic neutron scattering
  and density-functional methods}, Phys.\ Rev.\ B \textbf{68}, 144301 (2003).
\input{Schmalzl_prb_2003}

\bibitem{Elcombe_jpcss_1970}
M.~M. Elcombe and A.~W. Pryor, \emph{The lattice dynamics of calcium fluoride},
  J.\ Phys.\ C:\ Solid\ State\ Phys. \textbf{3}, 492 (1970).
\input{Elcombe_jpcss_1970}

\bibitem{Kaiser_pr_1962}
W.~Kaiser, W.~G. Spitzer, R.~H. Kaiser, and L.~E. Howarth, \emph{Infrared
  Properties of CaF$_2$, SrF$_2$, and BaF$_2$}, Phys.\ Rev. \textbf{127},
  1950--1954 (1962).
\input{Kaiser_pr_1962}

\bibitem{curtarolo:art93}
L.~A. Agapito, S.~Curtarolo, and M.~{Buongiorno~Nardelli}, \emph{Reformulation
  of $\mathrm{DFT}+U$ as a Pseudohybrid {H}ubbard Density Functional for
  Accelerated Materials Discovery}, Phys.\ Rev.\ X \textbf{5}, 011006 (2015).
\input{curtarolo:art93}

\bibitem{Andrade_Octopus_PCCP_2015}
X.~Andrade, D.~Strubbe, U.~{De Giovannini}, A.~H. Larsen, M.~J.~T. Oliveira,
  J.~Alberdi-Rodriguez, A.~Varas, I.~Theophilou, N.~Helbig, M.~J. Verstraete,
  L.~Stella, F.~Nogueira, A.~Aspuru-Guzik, A.~Castro, M.~A.~L. Marques, and
  A.~Rubio, \emph{Real-space grids and the {O}ctopus code as tools for the
  development of new simulation approaches for electronic systems}, Phys.\
  Chem.\ Chem.\ Phys. \textbf{17}, 31371--31396 (2015).
\input{Andrade_Octopus_PCCP_2015}

\bibitem{curtarolo:art116}
P.~D'{A}mico, L.~A. Agapito, A.~Catellani, A.~Ruini, S.~Curtarolo, M.~Fornari,
  M.~{Buongiorno~Nardelli}, and A.~Calzolari, \emph{Accurate \textit{ab initio}
  tight-binding Hamiltonians: Effective tools for electronic transport and
  optical spectroscopy from first principles}, Phys.\ Rev.\ B \textbf{94},
  165166 (2016).
\input{curtarolo:art116}

\bibitem{curtarolo:art86}
L.~A. Agapito, A.~Ferretti, A.~Calzolari, S.~Curtarolo, and
  M.~{Buongiorno~Nardelli}, \emph{Effective and accurate representation of
  extended Bloch states on finite Hilbert spaces}, Phys.\ Rev.\ B \textbf{88},
  165127 (2013).
\input{curtarolo:art86}

\bibitem{curtarolo:art111}
L.~A. Agapito, M.~Fornari, D.~Ceresoli, A.~Ferretti, S.~Curtarolo, and
  M.~{Buongiorno~Nardelli}, \emph{Accurate Tight-Binding Hamiltonians for 2{D}
  and Layered Materials}, Phys.\ Rev.\ B \textbf{93}, 125137 (2016).
\input{curtarolo:art111}

\bibitem{curtarolo:art108}
L.~A. Agapito, S.~Ismail-Beigi, S.~Curtarolo, M.~Fornari, and
  M.~{Buongiorno~Nardelli}, \emph{Accurate tight-binding Hamiltonian matrices
  from \textit{ab initio} calculations: Minimal basis sets}, Phys.\ Rev.\ B
  \textbf{93}, 035104 (2016).
\input{curtarolo:art108}

\bibitem{curtarolo:art129}
F.~Legrain, J.~Carrete, A.~{van~Roekeghem}, S.~Curtarolo, and N.~Mingo,
  \emph{How Chemical Composition Alone Can Predict Vibrational Free Energies
  and Entropies of Solids}, Chem.\ Mater. \textbf{29}, 6220--6227 (2017).
\input{curtarolo:art129}

\bibitem{aflowmlapi}
E.~Gossett, C.~Toher, C.~Oses, O.~Isayev, F.~Legrain, F.~Rose, E.~Zurek,
  J.~Carrete, N.~Mingo, A.~Tropsha, and S.~Curtarolo, \emph{{AFLOW-ML}: {A}
  {REST}ful {API} for machine-learning predictions of materials properties},
  submitted arxiv.org/1711.10744  (2017).
\input{aflowmlapi}

\bibitem{ICSD}
G.~Bergerhoff, R.~Hundt, R.~Sievers, and I.~D. Brown, \emph{The inorganic
  crystal structure data base}, J.\ Chem.\ Inf.\ Comput.\ Sci. \textbf{23},
  66--69 (1983).
\input{ICSD}

\bibitem{ICSD1}
V.~L. Karen and M.~Hellenbrandt, \emph{Inorganic crystal structure database:
  new developments}, Acta\ Cryst. \textbf{A58}, c367 (2002).
\input{ICSD1}

\bibitem{Liechtenstein1995}
A.~I. Liechtenstein, V.~I. Anisimov, and J.~Zaanen, \emph{Density-functional
  theory and strong interactions: Orbital ordering in {M}ott-{H}ubbard
  insulators}, Phys.\ Rev.\ B \textbf{52}, R5467--R5470 (1995).
\input{Liechtenstein1995}

\bibitem{Dudarev1998}
S.~L. Dudarev, G.~A. Botton, S.~Y. Savrasov, C.~J. Humphreys, and A.~P. Sutton,
  \emph{Electron-energy-loss spectra and the structural stability of {Ni}ckel
  oxide: An {LSDA}$+{U}$ study}, Phys.\ Rev.\ B \textbf{57}, 1505--1509 (1998).
\input{Dudarev1998}

\bibitem{curtarolo:art103}
P.~Gopal, M.~Fornari, S.~Curtarolo, L.~A. Agapito, L.~S.~I. Liyanage, and
  M.~{Buongiorno~Nardelli}, \emph{Improved predictions of the physical
  properties of {Zn}- and {Cd}-based wide band-gap semiconductors: A validation
  of the {ACBN0} functional}, Phys.\ Rev.\ B \textbf{91}, 245202 (2015).
\input{curtarolo:art103}

\end{thebibliography}
\end{document}